\documentclass[aps,pre,showpacs,floats,twocolumn,superscriptaddress,longbibliography,10pt]{revtex4-2}

\usepackage{comment}
\DeclareUnicodeCharacter{2009}{\,}
\UseRawInputEncoding

\usepackage{lipsum}
\usepackage{mathrsfs}
\usepackage{bm,amsbsy,amssymb,amsmath}
\usepackage{graphics,graphicx,dcolumn,epic,eepic,tabularx}

\usepackage{multirow,rotate,rotating,color}
\usepackage[utf8]{inputenc}
\usepackage{relsize}

\usepackage{xcolor}
\usepackage[utf8]{inputenc}

\definecolor{tuered}{RGB}{214,0,74}
\definecolor{tueblue}{RGB}{0,102,204}

\newcommand{\revisedtext}[1]{\textcolor{black}{#1}}

\usepackage{hyperref}

\usepackage{subcaption}

\begin{document}

\title{Self-assembled filament layers in drying sessile droplets: \texorpdfstring{\\}{ } from morphology to electrical conductivity}
\author{Johannes  Sch\"ottner}
\email{j.schoettner@fz-juelich.de}
\affiliation{Helmholtz Institute Erlangen-N\"urnberg for Renewable Energy (IET-2), Forschungszentrum J\"ulich, Cauerstra{\ss}e 1, 91058 Erlangen, Germany}

\author{Qingguang Xie}
\email{q.xie@fz-juelich.de}
\affiliation{Helmholtz Institute Erlangen-N\"urnberg for Renewable Energy (IET-2), Forschungszentrum J\"ulich, Cauerstra{\ss}e 1, 91058 Erlangen, Germany}

\author{Gaurav Nath}
\affiliation{Helmholtz Institute Erlangen-N\"urnberg for Renewable Energy (IET-2), Forschungszentrum J\"ulich, Cauerstra{\ss}e 1, 91058 Erlangen, Germany}

\author{Jens Harting}
\email{j.harting@fz-juelich.de}
\affiliation{Helmholtz Institute Erlangen-N\"urnberg for Renewable Energy (IET-2), Forschungszentrum J\"ulich, Cauerstra{\ss}e 1, 91058 Erlangen, Germany}
\affiliation{Department of Chemical and Biological Engineering and Department of Physics, Friedrich-Alexander-Universit\"at Erlangen-N\"urnberg, Cauerstra{\ss}e 1, 91058 Erlangen, Germany}

\date{\today}

\begin{abstract}

Controlling the deposition of filaments, such as nanowires and nanotubes, from evaporating droplets is critical for the performance of emerging technologies like flexible sensors and printed electronics. The final deposit morphology strongly governs functional properties, such as electrical conductivity, yet remains challenging to control. In this work, we numerically investigate how filament length, stiffness, and concentration affect deposition patterns during the drying process. We compare reaction-limited and diffusion-limited evaporation regimes, demonstrating that their distinct velocity fields and flow magnitudes fundamentally alter filament arrangement. While diffusion-limited evaporation drives the ``coffee-ring effect", compromising network uniformity, reaction-limited evaporation suppresses edge accumulation, promoting centered conductive deposits. We map out the spatial variation of filament alignment - tangential at the contact line, radial in the intermediate region, and random near the center. Longer filaments tend to favour more tangential alignment overall and suppress edge accumulation.
We find that by tuning the evaporation regime, filament deposition can lead to significantly lower percolation thresholds and significantly higher conductivity exponents.
These results quantify the link between evaporation kinetics and microstructure, providing guidelines for optimizing conductive network formation in printed electronics.

\end{abstract}

\maketitle

\section{Introduction}

Drying droplets are a versatile approach for assembling functional materials - such as nanowires, carbon nanotubes, and conducting polymers - into organized structures for a wide range of emerging applications. These include energy devices, sensors, neuromorphic circuits, biomedical systems, and printed electronics~\cite{Steinberger2024,Rao2022,Kwon2020,Wang2021}. In this process, functional nanomaterials are dispersed in liquid ink droplets and deposited onto a substrate via techniques like inkjet, aerosol-jet, or spray printing. As the solvent evaporates, internal convective flows, interfacial forces, and particle interactions drive the dispersed materials to self-organize into patterned, anisotropic, or hierarchical arrangements~\cite{Park2024, Dugyala2015}. Such evaporation-induced self-assembly enables the formation of tailored micro- and nanostructured films that are attractive for advanced manufacturing.

Device performance is strongly determined by the morphology of the deposited layer. In printed electronics, evaporation-induced self-assembly enables the fabrication of flexible and cost-effective filament layers~\cite{Kamyshny2019, Li2022}. For filament-based materials such as carbon nanotubes, silver nanowires, or conducting polymers, alignment reduces the number of filament crossings and enhances in-plane charge transport, while the quality of contact points and phase organization further improve conductivity and mechanical stability~\cite{Ren2012, Xu2020}. In addition to alignment, filament distribution and surface roughness critically affect electronic performance: rough or coiled deposits can create leakage pathways and limit efficiency, whereas embedding or planarization techniques reduce roughness and improve device reliability~\cite{HosseinzadehKhaligh2014}.

The final dried morphology of the deposit layer results from a complex interplay of fluid dynamics, interfacial effects, and particle interactions. The flows inside the droplets are mainly dependent on the evaporation modes (contact line dynamics) and evaporation regimes (evaporation rate/flux distribution). 
When a sessile droplet dries, the contact line typically follows one of three different modes: the constant contact angle mode (CCA, mobile contact line), the constant contact radius mode (CCR, pinned contact line), or a combination of these two, known as the stick-slide mode. In practice, printed droplets often exhibit stick-slide behavior, characterized by an initial period of pinning followed by depinning\revisedtext{~\cite{Wilson2023}}.
Evaporation regimes are generally classified as diffusion-limited or reaction-limited. Under diffusion-limited conditions, vapor transport away from the interface controls the evaporation rate. At low contact angles, the evaporation flux diverges near the contact line, generating strong outward capillary flows that promote ring-like deposits~\cite{Deegan1997}. By contrast, in reaction-limited (interface-limited) drying, interfacial kinetics govern the evaporation rate. The evaporation flux distribution along the droplet surface is more uniform, yielding weaker capillary flows and a more homogeneous deposition~\cite{Wilson2023, Larsson2023}. 
The morphology of the dried layer is also highly sensitive to parameters such as solvent composition or ambient conditions~\cite{Hu2006},
particle concentration~\cite{Baldwin2011}, aspect ratio~\cite{Dugyala2015} and stiffness~\cite{Park2024}. Depending on these factors, deposits can range from uniform coatings or aligned networks to uneven ring stains caused by the coffee-ring effect (CRE), where particles accumulate at the contact line.
Uniform deposits are often desirable, especially in coating processes and printing applications~\cite{Steinberger2024}.

The electrical conductivity of the deposited layer is one of the key properties for applications in printed electronics~\cite{Bi2024}. 
Percolation theory plays a key role in understanding and optimizing the conductivity of such morphologies, describing the transition from isolated clusters to a continuous, conductive network, governed by filament concentration, shape, and distribution~\cite{Berg2024}.
Filaments are particularly effective in achieving percolation at low concentrations, as their elongated shapes increase the likelihood of overlap and network formation. Their anisotropic geometry also promotes orientational ordering during deposition, which can enhance electrical, optical, and mechanical properties. This makes them attractive for applications requiring high conductivity and material efficiency~\cite{Cao2020}.

Experimental and theoretical approaches provide valuable insights, but face limitations in isolating and controlling the parameters of drying filament droplets. Numerical simulations overcome these constraints by enabling the systematic variation of individual parameters while keeping others fixed.
While spherical particles are well-studied~\cite{Xie2018, Xie2025, Hamjah2021},  
computational modeling of fluid-coupled rod-like structures forming deposition patterns during drying droplets remains largely unexplored~\cite{Kravchenko2020, Kolegov2025}.
To date, simulations of such systems have often been simplified, for instance by using Monte Carlo (MC) methods that treat filaments as random stick networks without incorporating the fluid dynamics of the evaporating droplet~\cite{Hicks2018, Fata2020}.

In this work, we couple the lattice Boltzmann method for fluid dynamics with a bead-spring approach for filaments.
Our approach enables the simulation of fluid flow, particle interactions, and network formation, providing insights into filament alignment, clustering, and percolation behavior following solvent drying. 
We consider the filament suspension in a dilute limit, where self-pinning is negligible, 
and focus on the stick–slide evaporation mode: droplets evaporate in CCR mode until a critical receding angle $\theta_{re}$ is reached, 
after which they transition to CCA mode~\cite{Orejon2011, Wilson2023}.
We investigate how filament length, concentration, stiffness, and evaporation conditions affect percolative network formation through a systematic parameter study. We then analyze the nematic order, radial density profiles, percolation probability, and electrical conductivity of the deposit.  

The remaining sections of this paper are organized as follows: Section II provides details of the simulation method, including the color-gradient multicomponent lattice Boltzmann (LB) model and the coupled bead-spring model for the filaments.  Section III presents results of drying filament-laden droplets and a discussion on deposit morphology, percolation and conductivity. Finally, Section IV summarizes our findings.

\section{Methods}
We simulate filaments suspended in an evaporating sessile droplet on a solid substrate. The fluids are described using the color-gradient lattice Boltzmann (CGLB) method, two-way coupled with a point-particle/bead-spring approach to resolve the motion and interactions of filaments.
\subsection{Color-gradient lattice Boltzmann model}
The fluid solver is based on the three-dimensional lattice Boltzmann method with $19$ discrete velocities (D3Q19)~\cite{Krger2017, Benzi1992}. 
We describe the fluid forming the droplet and the surrounding fluid as independent components. Each component $k \in \{1,2\}$ is described by a distribution function $f_i^k(\mathbf{x}, t)$, which evolves according to the lattice Boltzmann equation,
\begin{equation}
    f_i^k(\mathbf{x} + \mathbf{c}_i \Delta t, t + \Delta t) = f_i^k(\mathbf{x}, t) + \Omega_i^k(\mathbf{x}, t),
    \label{eq:LBE}
\end{equation}
where $\Omega_i^k$ is the collision operator, $\mathbf{c}_i$ are the discrete velocity vectors, and $i = 1, \dots, 19$. We set the timestep $\Delta t$ and the lattice constant $\Delta x$ to unity for simplicity.
The macroscopic density and velocity of component $k$ are given by
\begin{align}
    \rho_k(\mathbf{x}, t) &= \rho_0\sum_{i=1}^{19} f_i^k(\mathbf{x}, t), \label{eq:density} \\
    \mathbf{u}_k(\mathbf{x}, t) &= \frac{1}{\rho_k(\mathbf{x}, t)} \sum_{i=1}^{19} f_i^k(\mathbf{x}, t) \mathbf{c}_i, \label{eq:velocity}
\end{align}
\revisedtext{with $\rho_0 = 1$ as the reference density. The unit of mass is defined as $m_0 = \rho_0\, \Delta x^3 = 1$, and the unit of energy as $E_0 = m_0 ( \Delta x / \Delta t )^2 = 1$.}

The CGLB method, enables the simulation of immiscible or partially miscible fluids by modeling interfacial tension and enforcing phase segregation through three substeps: relaxation, perturbation, and recoloring~\cite{Gunstensen1991, Leclaire2017, Aursj2015}.
The total collision operator $\Omega_i$ acting on the total distribution function $f_i$ is decomposed as
\begin{equation}
    f_i^*(\mathbf{x}, t) = (\Omega_i^{\text{recol}} \circ \Omega_i^{\text{pert}} \circ \Omega_i^{\text{BGK}})[f_i(\mathbf{x}, t)],
    \label{eq:collision}
\end{equation}
where $f_i(\mathbf{x}, t) = f_i^1(\mathbf{x}, t) + f_i^2(\mathbf{x}, t)$ is called as the color-blind distribution function. 

\textit{Relaxation: }The Bhatnagar–Gross–Krook (BGK)
operator relaxes $f_i$ toward equilibrium:
\begin{equation}
    \Omega_i^{\text{BGK}}[f_i] = f_i - \frac{1}{\tau} \left( f_i - f_i^{\text{eq}} \right)
    \label{eq:BGK}
\end{equation}
\revisedtext{Here, $\tau$ is the dimensionless relaxation time. Through the Chapman–Enskog expansion, $\tau$ is related to the kinematic viscosity $\nu = c_s^2(\tau - 0.5)$, with $c_s=\frac{1}{\sqrt{3}}\frac{\Delta x}{\Delta t}$ the speed of sound for the D3Q19 lattice.
The equilibrium distribution is given by}
\begin{equation}
    f_i^{\text{eq}} = \rho\, w_i \left( 1 + \frac{\mathbf{c}_i \cdot \mathbf{u}}{c_s^2} + \frac{(\mathbf{c}_i \cdot \mathbf{u})^2}{2c_s^4} - \frac{\mathbf{u}^2}{2c_s^2} \right),
    \label{eq:equilibrium}
\end{equation}
\revisedtext{where $\rho = \rho_1 + \rho_2$ is the total density and $\mathbf{u} = \frac{1}{\rho} \sum_{i} f_i \mathbf{c}_i$ is the mixture velocity~\cite{Krger2017}.} The lattice weights $w_i$ for the D3Q19 lattice are
\begin{equation}
w_i = \begin{cases} 
    1/3, & i = 1 \\
    1/18, & i = 2,\dots,7 \\
    1/36, & i = 8,\dots,19
\end{cases}.
\end{equation}

\textit{Perturbation:} Interfacial tension is introduced via a perturbation proportional to the color gradient $\mathbf{G}$ as
\begin{equation}
    \Omega_i^{\text{pert}} [f_i] = f_i + \frac{A}{2} |\mathbf{G}| \left( w_i \cos^2 \phi_i - B_i \right),
    \label{eq:perturbation}
\end{equation}
where $\phi_i$ is the angle between $\mathbf{G}$ and $\mathbf{c}_i$. The parameter $A$ sets the surface tension magnitude $\sigma$ according to $A=\frac{9\sigma}{4\tau}$~\cite{Leclaire2017}, while the lattice-dependent weights $B_i$ are defined as
\begin{equation}
B_i = \begin{cases}
    -2/9, & i = 1 \\
    1/54, & i = 2,\dots,7 \\
    1/27, & i = 8,\dots,19
\end{cases}.
\end{equation}
\revisedtext{We note that the unit of surface tension is $E_0 / \Delta x^2 = 1$ in the simulations}.

The normalized color field $\phi$ is defined as
\begin{equation}
\phi = \frac{\rho_1 - \rho_2}{\rho_1 + \rho_2},
\end{equation}
and the color gradient is approximated by
\begin{equation}
    \mathbf{G}(\mathbf{x}, t) = \nabla \phi(\mathbf{x}, t) \approx \sum_{i} w_i \mathbf{c}_i\, \phi(\mathbf{x} + \mathbf{c}_i \Delta t, t).
    \label{eq:gradient_discretization}
\end{equation}

\textit{Recoloring }enforces immiscibility by redistributing $f_i$ to $f_i^k$,
\begin{equation}
    f_i^k = \frac{\rho_k}{\rho} f_i + \beta \frac{\rho_1 \rho_2}{\rho^2} \cos \phi_i\, f_i^{\text{eq}}(\rho_k, \mathbf{0}),
    \label{eq:recoloring}
\end{equation}
where $\beta = 0.99$ controls interface sharpness and $f_i^{\text{eq}}(\rho_k, \mathbf{0})$ is the equilibrium distribution for component $k$ at zero velocity.
The updated component distributions $f_i^k$ are then streamed to complete the LB time step.

\revisedtext{Solid boundaries are treated using the half-way bounce-back boundary condition, enforcing a no-slip condition with the effective wall located halfway between fluid and solid lattice nodes~\cite{Krger2017}.}
To describe wetting on the substrate, we follow the work of Akai et al.~\cite{Akai2018}. We dynamically adjust the fluid interface normal $ \mathbf{n}_f $ at the triple line to enforce the contact angle condition. The algorithm computes two candidate normals $ \mathbf{v}_0 $ and $ \mathbf{v}_1 $ based on the geometric relationship between the fluid interface normal $ \mathbf{n}_f $ and the wall normal $ \mathbf{n}_w $:
{\small
\begin{eqnarray}
     \mathbf{v}_{0} &=& \left( \cos(\theta_f) - \frac{\sin(\theta_f)}{\sin(\theta_i)} \cos(\theta_i) \right) \mathbf{n}_w + \frac{\sin(\theta_f)}{\sin(\theta_i)} \mathbf{n}_f  \\
    \mathbf{v}_{1} &=& \left( \cos(\theta_f) + \frac{\sin(\theta_f)}{\sin(\theta_i)} \cos(\theta_i) \right) \mathbf{n}_w -\frac{\sin(\theta_f)}{\sin(\theta_i)} \mathbf{n}_f
\end{eqnarray}}
Here, $ \theta_f $ is the target contact angle, $ \theta_i = \cos^{-1}(\mathbf{n}_f \cdot \mathbf{n}_w) $ is the initial contact angle, and $ \mathbf{n}_w $ is the wall normal. The final fluid interface normal $ \mathbf{n}_f $ is updated to the candidate ($ \mathbf{v}_0 $ or $ \mathbf{v}_1 $) that minimizes the deviation from the target contact angle. This approach ensures accurate enforcement of the dynamic contact angle condition during droplet spreading or receding.

\subsection{Evaporation model}\label{Evap_model}
Fluid evaporation is implemented following the recent work of Nath et al.~\cite{Nath2025}.
This reaction-limited evaporation model simulates phase change by converting distribution populations at the liquid-vapor interface. This approach is adaptable to any evaporation regime with a known analytical evaporation flux expression and is computationally efficient, as calculations are confined to the interface. 
Evaporation is modeled by converting resting populations at \revisedtext{lattice sites at the interface, identified by a color-gradient magnitude larger than a threshold value $\Gamma=0.305$~\cite{Nath2025}:
\begin{equation}\label{S_factor}
\begin{aligned}
    f_0^{1}(\mathbf{x}, t)^{\text{new}} &= f_0^{1}(\mathbf{x}, t) - \varphi\,\Delta t, \\
    f_0^{2}(\mathbf{x}, t)^{\text{new}} &= f_0^{2}(\mathbf{x}, t) + \varphi\,\Delta t .
\end{aligned}
\end{equation}}
The surface mass flux $\Xi$ across each site is linked to a volumetric sink rate $\varphi = \frac{d\rho}{dt} =  S\Xi$, with a correction factor $S:=1/3$ accounting for discretization errors~\cite{Nath2025}. During each timestep, mass is subtracted from one fluid component and added to another, preserving total mass and momentum.

In printing technologies, evaporation typically occurs in an intermediate regime between the diffusion-limited and reaction-limited limits, which can be adjusted by masking the droplet, varying the relative humidity\revisedtext{,} the ambient pressure\revisedtext{,} by imposing an airflow~\cite{Majewski2025} \revisedtext{or when molecule transfer through the interface is hindered by the solutes}. \revisedtext{We restrict our analysis to the dilute regime, assuming negligible particle adsorption at the interface, allowing us to neglect evaporation shielding effects caused by filament accumulation.}

In the \textit{reaction-limited} regime, interfacial kinetics - rather than vapor diffusion - sets the evaporation rate. This commonly occurs on heated substrates or at low pressure~\cite{MURISIC2011,Nath2025}. The surface mass flux $\Xi$ follows the Hertz–Knudsen (HK) relation, \revisedtext{which provides the kinetic boundary condition linking the interfacial mass flux to the local thermodynamic state of the interface}. Using the Clausius-Clapeyron law, the \revisedtext{non-dimensional} HK relation is given by
\begin{equation}
    \Xi = \frac{\Lambda \, \rho_v \, \mathcal{L}}{T_{\text{sat}}^{3/2}} 
    \sqrt{\frac{M_v}{2\pi R_g}} \, \left[\revisedtext{T_I} - T_{\text{sat}}\right],
    \label{eq:HK}
\end{equation}
with $\Lambda$ the accommodation coefficient, $\rho_v$ the saturation vapor density, $M_v$ the molecular mass of the vapor, $R_g$ the universal gas constant, $\mathcal{L}$ the latent heat of vaporization, \revisedtext{$T_I$} the liquid–vapor interface temperature, and $T_{\text{sat}}$ the saturation temperature.
In the non-equilibrium one-sided model (NEOS)~\cite{Larsson2023, MURISIC2011}, under the thin-droplet assumption, with negligible substrate thickness and curvature effects, \revisedtext{the interface temperature is expressed in terms of the local film height and inserted into the HK relation, yielding the evaporation flux $J$ for thin droplets} 
\begin{equation}\label{Reaction_limited}
  J \;=\; \frac{J_0}{K + \tilde{h}}.
\end{equation}
Here, $\tilde{h}$ is the droplet height, normalized by the initial height, $K$ is a dimensionless kinetic resistance (the inverse mass-transfer coefficient, typically $K \approx 10$ for water), and $J_0$ represents a characteristic flux set by the thermal driving force.

In the case of \textit{diffusion-limited} evaporation of a sessile droplet, vapor transport is assumed quasi-steady and purely diffusive, governed by the Laplace equation~\cite{Deegan1997, Hu2002}
\begin{equation}
\nabla \cdot (D_v \nabla u) = 0.
\end{equation}
Here, $D_v$ is the vapor diffusivity in air and $u(\mathbf{x})$ the vapor concentration field. The boundary conditions are $ u = u_s $ (saturated vapor concentration) at the droplet surface and $ u = u_\infty $ (ambient vapor concentration) far away. The droplet is modeled as a spherical cap (small Bond number), and viscous forces are negligible (low capillary number).
The resulting evaporation flux $ J(r) $ exhibits a divergence near the contact line. Nevertheless, the total evaporation rate remains finite because the singularity is integrable; in addition, real droplets often possess a thin precursor film that regularizes the divergence~\cite{Saxton2016, OBrien2024}. For small contact angles $ (\theta \to 0) $, the evaporation flux is written as
\begin{equation}
J(r) \propto \frac{1}{\sqrt{a^2 - r^2}},
\end{equation}
where $a$ is the droplet base radius and $r$ the radial position (see Fig.~\ref{fig:illustration1}).
For arbitrary angles, an approximate form is~\cite{Hu2002}
\begin{equation}
J \propto (a^2 - r^2)^{-\lambda}, \quad \lambda = 0.5-\theta/\pi,
\end{equation}
in which $\theta(t) = 2 \arctan\left( \frac{h_0(t)}{a} \right)$ for a spherical droplet in case of semi-steady evaporation.
Furthermore, the evaporation flux can be approximated as~\cite{MURISIC2011}
\begin{equation}
    J \approx \frac{J_0}{\tilde{h}^{\Psi}}, \quad \Psi \in [1,2],
\end{equation}
with $\Psi \approx 1$ for small contact angles. Notably, in this regime, the diffusion-limited evaporation profile closely resembles that of reaction-limited evaporation for high-volatility liquids (i.e., small evaporation resistance or low $K$-values). In the following, we investigate the diffusion-limited evaporation regime ($K=0.031\approx0$) and the reaction-limited evaporation regime ($K=10$) with the evaporation flux given in Eq.~\ref{Reaction_limited}.

In the diffusion-limited regime, the vapor fields of neighboring droplets overlap, leading to a \textit{shielding effect} that reduces the local evaporation rate. \revisedtext{For two identical adjacent droplets} the resulting \textit{local evaporative flux} is expressed as~\cite{Wray2021}
\begin{equation}
    J(r, \xi) = J(r) \left[ 1 - \frac{M\sqrt{b^2 - a^2}}{2\pi \left(r^2 + b^2 - 2rb \cos \xi\right)} \right],
    \label{eq:local_flux}
\end{equation}
where $r$ is the radial coordinate from the center, $\xi$ the azimuthal angle measured from the axis connecting the droplet centers, \revisedtext{$a$ is the base radius, and $b$ the center-center-distance between the droplets. The shielding factor $M$ quantifies the reduction of the evaporation flux due to the neighboring droplet and takes the analytical form~\cite{Wray2021}}
\begin{equation}
    M = \frac{4}{1 + \frac{2}{\pi} \arcsin\!\left(\frac{a}{b}\right)}.
    \label{eq:shielding_factor}
\end{equation}

\subsection{Filament model}\label{S:Polymer}
The filaments within the droplet are modeled using a bead-spring representation. Each filament is constructed as a sequence of connected beads (monomers) linked by springs.
Inter-monomer bonds are governed by the finite extensible nonlinear elastic (FENE) potential, known to accurately model real filament behavior~\cite{Li2012}:
\begin{equation}
    U_{\text{FENE}}(d) = -\frac{1}{2} k_s R_{m}^2 \ln\left( 1 - \frac{d^2}{R_{m}^2} \right), \quad d < R_{m}
\end{equation}
Here, $ d $ is the separation between adjacent beads, $ k_s=0.3 $ is the spring constant, and $ R_{m}=1.2 $ is the maximum bond extension. 
To incorporate excluded volume effects, a \revisedtext{WCA-like (Weeks-Chandler-Andersen)} potential 
\begin{equation}
    U_{\text{WCA}}(d) = \varepsilon \left[ \left( \frac{d_0}{d} \right)^{12} - 2 \left( \frac{d_0}{d} \right)^6 \right] + \varepsilon, \quad d \leq d_0
\end{equation}
is added, where $ \varepsilon=0.03 $ is the depth of the potential well, and $d_0=1$ is the characteristic particle diameter (corresponding to the distance at which the potential vanishes). \revisedtext{The} stiffness of filaments is modeled using an approximate harmonic potential for second-nearest-neighbor interactions,
\begin{equation}
    U_{\text{H}}(d) \approx \frac{1}{2} k (d - 2d_0)^2,
\end{equation}
where $ k=8.0 $ is the spring constant and $ 2d_0 $ is the equilibrium distance between second-nearest neighbors.

Combining WCA, FENE, and second-nearest-neighbor potentials ensures that each spring resists both overextension and overlap, thereby maintaining the filament’s elasticity and excluded-volume stability. 
The harmonic term governs the chain’s resistance to bending, capturing the physical flexibility characteristic of polymers or nanowires in the fluid.

Filaments are two-way coupled to the LB fluid via a local momentum-exchange scheme. Each bead experiences a hydrodynamic drag force~\cite{Ahlrichs1999,Jung2022}
\begin{equation}
    \mathbf{F}_{\text{drag}} = -\gamma (\mathbf{v}_{\text{p}} - \mathbf{v}_{\text{f}}),
\end{equation}
with $\gamma = 6\pi\eta a_0$ representing the friction coefficient for a sphere of radius $a_0$ in a fluid of dynamic viscosity $\eta$. 
Here, $\mathbf{v}_{\text{p}}$ denotes the bead velocity, while $\mathbf{v}_{\text{f}}$ is the fluid velocity interpolated at the bead’s position.

Conversely, the force exerted by the bead on the fluid is mapped back to the lattice using the method proposed by Ahlrichs and Dünweg~\cite{Ahlrichs1999}. This scheme distributes the momentum to the nearest lattice nodes via a regularized delta function. The update rule for the component distributions $f_i^k$ is given by
\begin{equation}
    f^k_i(\mathbf{x}) \leftarrow f^k_i(\mathbf{x}) - \frac{w_i \rho_k(\mathbf{x}, t)}{c_s^2} \left( \mathbf{F}_{\text{drag}} \cdot \mathbf{c}_i \right) \delta(\mathbf{x}, \mathbf{x}_{\text{p}}).
\end{equation}
\revisedtext{The weight function $\delta(\mathbf{x}, \mathbf{x}_{\text{p}})$ corresponds to a trilinear interpolation kernel~\cite{Ahlrichs1999}, assigning linearly weighted contributions to the nearest lattice nodes in each spatial direction.}
This formulation locally injects momentum into the fluid while ensuring conservation.

The solvation force, introduced by Sega et al.~\cite{Sega2013}, is used to confine filaments inside the droplet. It depends on the local gradients of the fluid-component densities,
\begin{equation}
   \mathbf{F}_{\text{sol}}^i
   = - \sum_k \kappa_k \nabla \rho_k(\mathbf{r}_i),
\end{equation}
where $\rho_k$ denotes the density field of fluid component $k$, and $\kappa_k$ controls the strength and sign of the interaction: $\kappa_k > 0$ attracts particles toward regions of higher $\rho_k$, while $\kappa_k < 0$ leads to repulsion from dense regions.

Since the system is initialized in a dilute regime, feedback from particle density gradients on the fluid can be neglected. At later times, when particles accumulate near the substrate, this back-coupling remains negligible because particles become immobilized 
by static friction at the substrate, and no longer affect the surrounding flow.

To capture capillary interactions, we calibrate the solvation magnitude $\kappa_k$ by balancing the net inward solvation force on the wetted segment $L_w$ against the capillary compression. Following the formulation established by Seong et al.~\cite{Seong2017}, we approximate the theoretical capillary force $F_c$ as
\begin{equation}
F_c \approx 0.684\,\frac{2\sigma d_0 L_w}{R_0}.
\end{equation}
The prefactor $0.684$ is a geometric scaling factor adopted from~\cite{Seong2017}.

The friction between filaments and the substrate critically influences the final deposition pattern during droplet evaporation. 
For adhesive surfaces, Derjaguin extended Amontons' law by including an adhesion-related term~\cite{Derjaguin1934}. 
Accordingly, the friction force can be decomposed into external and internal contributions as~\cite{Gao2004}
\begin{equation}
    F_{\text{friction}} = \mu (P_0 + P) = \tau_{\!s} A + \mu P,
\end{equation}
where $\mu$ is the friction coefficient, $P$ the external load, $P_0$ the internal (adhesive) load, $A$ the real contact area, and $\tau_{\!s}$ the interfacial shear strength, i.e., the shear stress that adhesive junctions can sustain before sliding.
This explains the existence of a finite friction force even at zero external load due to adhesion.
The internal component originates from van der Waals forces, commonly modeled by the Lennard-Jones potential
\begin{equation}
    U_{\text{LJ}}(d) = \varepsilon \left[ \left( \frac{d_0}{d} \right)^{12} - 2 \left( \frac{d_0}{d} \right)^6 \right]
\end{equation}
where $ d $ is the distance from the substrate, and $ \varepsilon=1.5 \times 10^{-3} $ the potential well depth. This interaction restricts bead mobility, effectively generating internal friction.
As evaporation proceeds, frictional forces can exceed solvation forces, immobilizing filaments and producing ring-like deposition patterns~\cite{Xie2018}.

The equations of motion of the filament beads and their interactions are integrated using the standard leap-frog algorithm, updating positions at integer and velocities at half-integer time steps.

\subsection{Problem definition and assumptions}
We investigate how filament length, concentration, stiffness, and evaporation modes influence deposition patterns. Simulations are performed on a lattice of size $96 \times 96 \times 48$ nodes, initialized with an equilibrated hemispherical droplet of base radius $h_0=a=38$ nodes \revisedtext{corresponding to a droplet radius of approximately $2\mu m$~\cite{Waele2025}}, as shown in Fig.~\ref{fig:illustration1}. \revisedtext{This scale represents water with dynamic viscosity \(\eta_w = 1 \times 10^{-3}\,\mathrm{Pa \cdot s}\), density \(\rho_w = 10^{3}\,\mathrm{kg/m^3}\), and surface tension \(\sigma_w = 7.2 \times 10^{-2}\,\mathrm{N/m}\). Filaments have a width of  \(\approx 50\,\mathrm{nm}\) and a maximum length of  \(\approx 1.5\,\mu m\).}
The evaporation rate is chosen so that complete evaporation for single droplets occurs over $N_t = 200{,}000$ simulation steps, defining a characteristic interface velocity $v_c = a/N_t = 1.9 \times 10^{-4}$. The surface tension $\sigma=2 \times 10^{-3}$ is chosen as high as possible so that the droplet has the shape of a spherical cap, but small enough to minimize spurious currents without inducing numerical instability in the filament acceleration integration. We implement a stick-slide evaporation mode: the contact line remains pinned until the contact angle decreases to approximately $40^\circ$, after which retraction occurs\revisedtext{~\cite{Nguyen2012}}. \revisedtext{The evaporation flux scale $J(0)$ is chosen to reflect realistic printing conditions, ensuring that key dimensionless numbers in the simulations match experimental values.}
The simulation operates in a capillary-dominated regime where gravity is negligible, justified by a small Bond number ($\mathrm{Bo} \ll 1$). We assume isothermal conditions with surfactant-free droplets; consequently, Marangoni flows are neglected ($\mathrm{Ma} \ll 1$)~\cite{Greyson2017,Hu2006}. 
The flow is characterized by low Reynolds numbers ($\mathrm{Re} \ll 1$), indicating that viscous forces dampen inertial effects.
Solute transport is convection-dominated ($\mathrm{Pe} \gg 1$), with Brownian diffusion suppressed by chain connectivity~\cite{Winkler2007, Wang2021_2}. To ensure stable fluid-filament coupling, we utilize a bead mass of $m = 60$, resulting in a low Stokes number ($\mathrm{St} \ll 1$). This confirms that the inertial response of the beads is negligible compared to viscous damping. Consequently, the beads effectively trace the local fluid velocity, subject only to confinement by the receding interface and interactions with other beads and the substrate. \revisedtext{This is consistent with overdamped filament dynamics in typical printing scenarios}.

\section{Results and Discussion}

\subsection{Evaporation of a pure droplet}
We first examine the evaporation of a pure droplet and validate the evaporation model by comparing the time evolution of the droplet height $h(r,t)$ and the radial capillary flow $v(r,t)$ with theoretical predictions.
The so-called \textit{rush-hour} behavior, characterized by an increase in $v(r,t)$ over time, is expected in the CCR mode~\cite{Marn2011}, but is partially suppressed here due to deviations from the ideal spherical-cap shape and dynamic pinning~\cite{Li2016, Zhang2015}. Once the receding contact angle of $\theta\approx40^\circ$ is reached, the capillary flow driven by pinning diminishes in the CCA mode~\cite{Greyson2017}. Our simulations presented below extend previously published benchmarks that were limited to small contact angles in the CCR mode and uniform evaporation~\cite{Xie2018, Nath2025}. \revisedtext{While the current study focuses on a dilute filament concentration, we note that the presence of filaments increases the effective viscosity of the suspension~\cite{Kol2021} and correspondingly suppresses capillary flows.}

\begin{figure}
    \centering
    \includegraphics[width=1.0\linewidth]{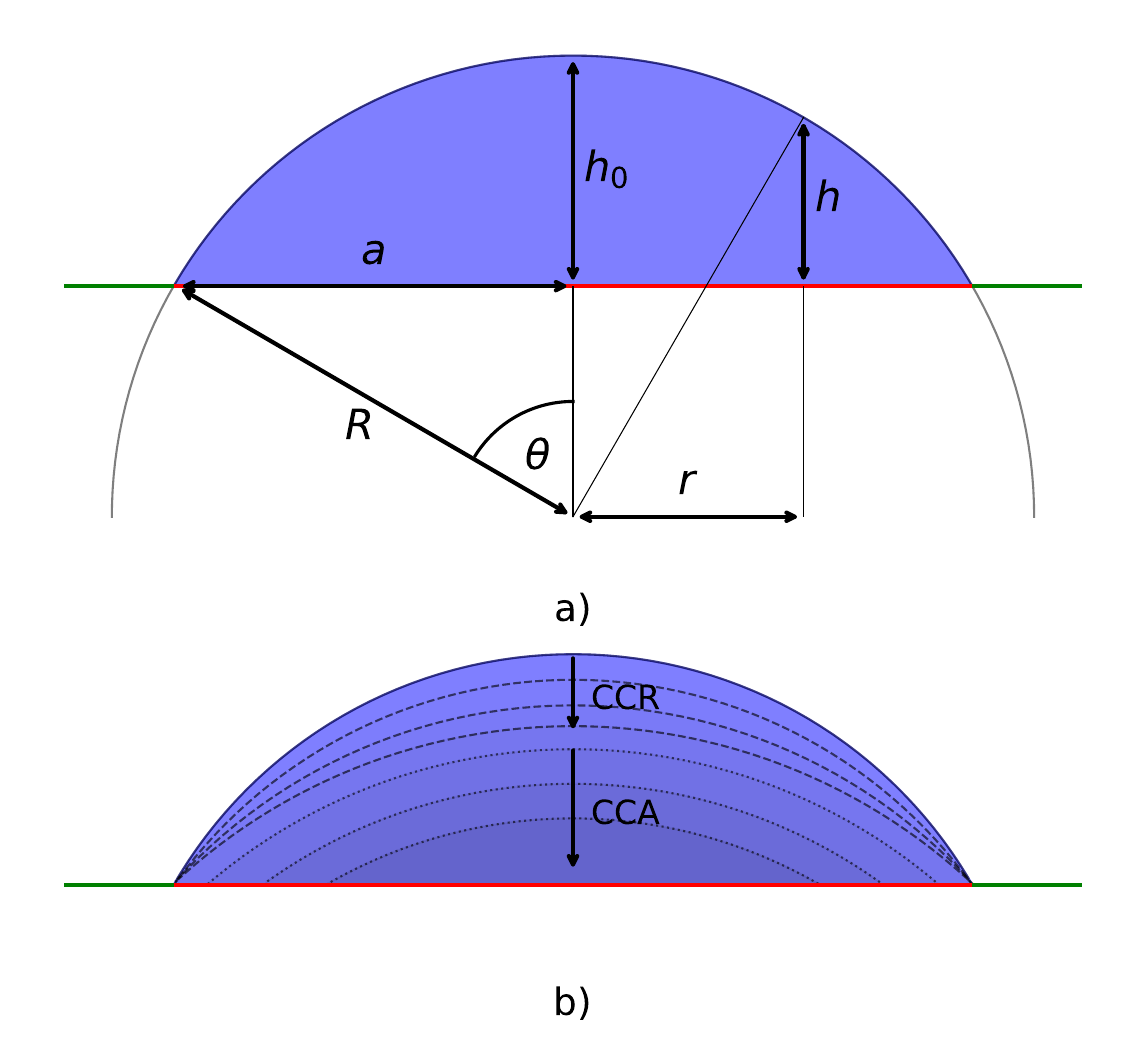}
    \caption{\revisedtext{A resting droplet on a patterned substrate with variable wettability: a central circular hydrophilic region (red, $\theta \approx 40^\circ$) surrounded by a neutral wetting area (green, $\theta=90^\circ$). (a) Geometry of a spherical cap droplet with radius of curvature $R$, base radius $a$, height $h_0$, and local height $h$ at radial position $r$. (b) Stick-slip dynamics illustrated with later time steps shown in darker colors: the dashed line marks the constant contact radius mode (CCR), followed by dotted lines indicating the constant contact angle mode (CCA).}}

    \label{fig:illustration1}
\end{figure}

Fig.~\ref{fig:illustration1} illustrates the geometry of an axisymmetric spherical-cap droplet in cylindrical coordinates $(r,z)$. Its volume $V$, radius of curvature $R$, and contact angle $\theta$ are given by
\begin{eqnarray}
V &=& \frac{\pi}{6}h_0(3a^2 + h_0^2), \label{Volume_SC}\\
R &=& \frac{a^2 + h_0^2}{2h_0},\\
\quad \theta &=& \arcsin\!\left(\frac{2h_0a}{a^2 + h_0^2}\right).
\end{eqnarray}
The height profile $h(r,t)$ is
\begin{equation}
\begin{split}
    h &= \sqrt{R^2 - r^2} - R\cos{\theta} \\
        &= \sqrt{\left(\frac{h_0^2 + a^2}{2h_0}\right)^2 - r^2} 
           - \frac{a^2 - h_0^2}{2h_0},
\end{split}
\end{equation}
which, for small contact angles, simplifies to~\cite{Marn2011_2}
\begin{equation}\label{rev1}
h \approx h_0 \frac{a^2 - r^2}{a^2 + h_0^2}.
\end{equation}

\subsubsection{Mass conservation}
Mass conservation dictates that for perfect contact line pinning, the rate of change of liquid height in an infinitesimal annular element at radius $r$ equals the net radial flux into the element minus the evaporated mass~\cite{Deegan2000, Marn2011_2}:
\begin{equation}\label{GoverningEq}
\frac{\partial h}{\partial t} 
= -\frac{1}{r} \frac{\partial \revisedtext{Q}}{\partial r} 
  - \frac{1}{\rho} J \sqrt{1 + \left(\frac{\partial h}{\partial r}\right)^2}
\end{equation}
Here, $Q=r h v$ is the radial volume flux, $\rho$ is the liquid density, $J$ the local evaporative flux. 
The geometrical factor $\sqrt{1 + (\partial_r h)^2}$ accounts for the local surface inclination. 

Eq.~\ref{GoverningEq} describes the height evolution of a spherical-cap droplet during evaporation in the CCR regime. 
To extend it to imperfect pinning, we account for the change in the droplet height $h_0(t)$ caused by variations in the base radius $a(t)$ under volume conservation. 
\begin{equation}
V = \frac{\pi}{6} h_0 \left( 3a^2 + h_0^2 \right), 
\qquad \frac{\partial V}{\partial t} = 0.
\end{equation}
Differentiating gives
\begin{equation}
\begin{split}
\frac{\partial V}{\partial t} &= \frac{\pi}{6} \left[ 
\frac{\partial h_0}{\partial t} \left( 3a^2 + h_0^2 \right) \right. \\
&\quad \left. +\, h_0 \left( 6a \frac{\partial a}{\partial t} + 2h_0 \frac{\partial h_0}{\partial t} \right) \right] = 0.
\end{split}
\end{equation}
Simplifying and solving for $\frac{\partial h_0}{\partial t}$ yields
\begin{equation}\label{equation_1}
\frac{\partial h_0}{\partial t} 
= - \frac{2 a h_0}{a^2 + h_0^2} \frac{\partial a}{\partial t}.
\end{equation}
The time derivative of the local droplet height $h(r,t)$ follows from the chain rule \revisedtext{and Eq.~\ref{rev1}} as 
\begin{equation}\label{ConservedVolume}
\begin{split}
\zeta_r &:= \left. \frac{\partial h}{\partial t} \right|_V 
= 
\frac{\partial h}{\partial a} \frac{\partial a}{\partial t} 
+ \frac{\partial h}{\partial h_0} \frac{\partial h_0}{\partial t} \\
&= \frac{2a h_0 (h_0^2 + r^2) + 4a h_0^3 (a^2 - r^2)}
{(a^2 + h_0^2)^2} \frac{\partial a}{\partial t},
\end{split}
\end{equation}
This contribution is negligible under perfect pinning ($\frac{\partial a}{\partial t}=0$) but dominates at small contact angles during depinning. \revisedtext{Including this correction term, the governing equation becomes}
\begin{equation}\label{GoverningEq2}
\revisedtext{\frac{\partial h}{\partial t} 
= -\frac{1}{r} \frac{\partial \revisedtext{Q}}{\partial r} 
  - \frac{1}{\rho} J \sqrt{1 + \left(\frac{\partial h}{\partial r}\right)^2} - \zeta_r.}
\end{equation}

\subsubsection{Droplet Height}
\begin{figure}
    \centering
    \includegraphics[width=1.0\linewidth]{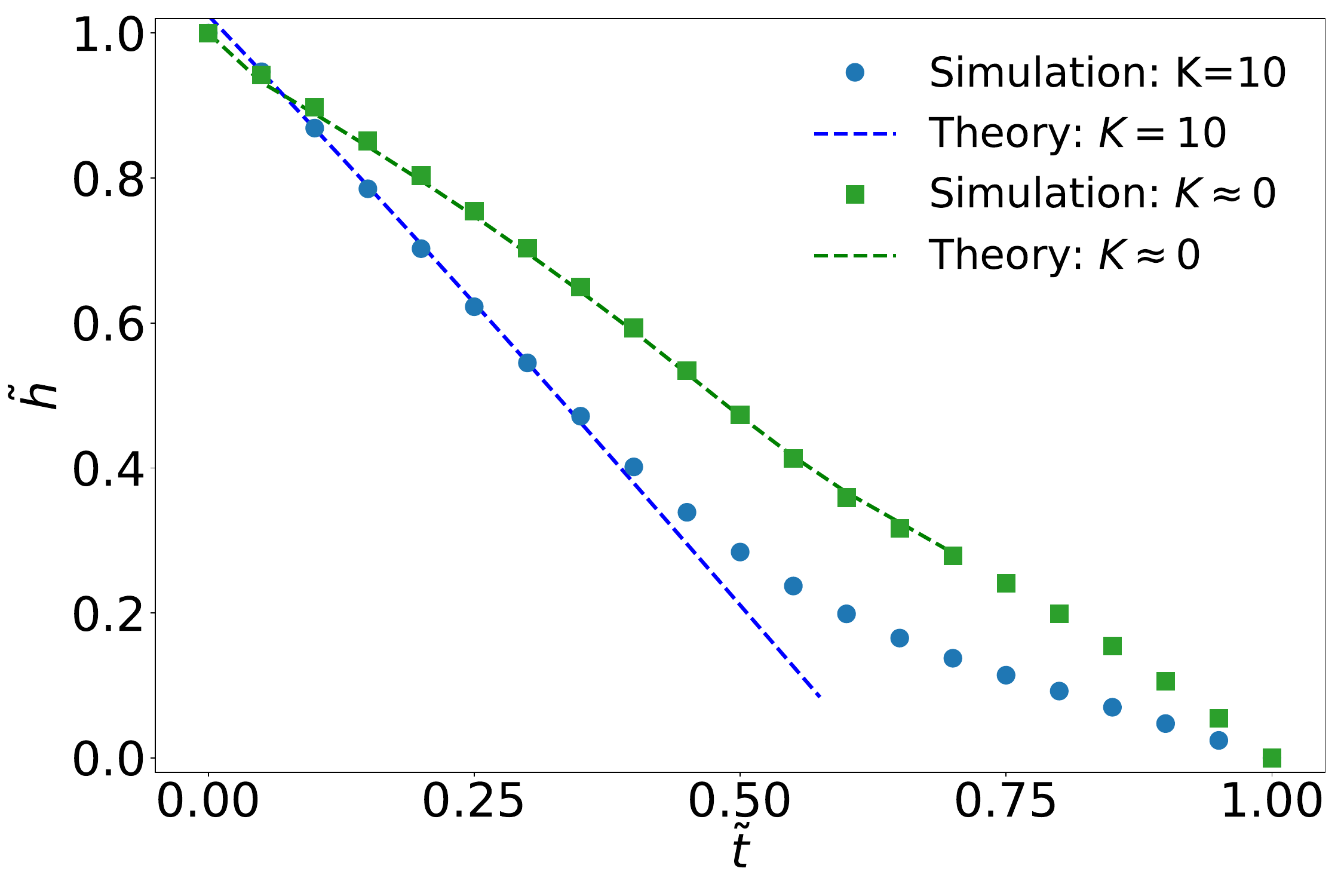}
    \caption{Time evolution $\tilde{t} =\frac{t}{N_t}$ of the relative droplet height $\tilde{h}=h/h_0$ at the droplet center ($r=0$) during stick-slide evaporation. For $K=10$, simulation (blue circles) follows Eq.~\ref{Prediciton_height} (blue line) during the pinned (CCR) stage; for $K \approx 0$, simulation (green squares) agrees with the theoretical prediction (green line), obtained by numerically solving Eq.~\ref{Volume_SC} and Eq.~\ref{change_vol}. 
    At late stages ($\tilde{t} \gtrsim 0.6$), when the droplet height falls below roughly 10 lattice units, deviations appear due to the CGLB discretization limits.
    }
    \label{fig:Theory_Height}
\end{figure}

The droplet height decreases over time due to evaporation, with the rate of volume loss given by
\begin{equation}\label{change_vol}
    \frac{dV}{dt} = -\frac{1}{\rho}\int_A J(r)\, dA,
\end{equation}
where $J(r)$ is the local evaporative flux integrated over the spherical-cap surface $A$.
Fig.~\ref{fig:Theory_Height} shows the droplet height $\tilde{h}(t)=h(t)/h_0$ as a function of time during evaporation, compared with theoretical predictions in both the reaction-limited ($K=10$) and diffusion-limited ($K\approx 0)$ regimes.

In the reaction-limited regime ($K=10$) with nearly uniform evaporation, an analytical expression can be obtained. For a droplet in the CCR mode, the height is given by~\cite{Nath2025}
\begin{equation} \label{Prediciton_height}
h_{\mathrm{CCR}}(t)
= h_0 - \frac{2\langle J \rangle}{\rho} t,
\end{equation}
where $\langle J \rangle$ represents the spatially averaged flux. \revisedtext{Here, we use $h(t)$ to denote $h_{r=0}(t)$.} The total evaporation time is defined as
$t_e = h_0 \rho / (2\langle J \rangle)$.
For $J_0 = 0.0005$, \revisedtext{the slope obtained from linear regression differs from the theoretical prediction by less than $0.3\%$,} confirming that the model reproduces the expected linear decay of the droplet during the CCR stage.
Adjusting the threshold parameter controlling the interface width may further improve the agreement~\cite{Nath2025}.

In the diffusion-limited regime ($K\approx0)$, the non-uniform flux requires numerical integration of $J(r)$ to compute the height evolution. For small droplets at late stages ($h \lesssim 10$ lattice units), deviations appear due to the CGLB discretization limits \revisedtext{(the interface is diffusive with a typical thickness of around 6 lattice units)}, with the theory predicting a slightly faster evaporation. This discrepancy is negligible for our analysis, as inertia is minimal and most particles are deposited by this stage. Applying a late-time linear fit to extract a correction factor $S$ (see Eq.~\ref{S_factor}) restores consistency with our simulations. \revisedtext{Due to the interface thickness, we neglect detailed filament deposition and growth dynamics at the contact line. For droplets with a radius comparable to or smaller than the filament length, this approximation holds well; however, for larger droplet-to-filament ratios, the spatial distribution of deposits and alignment patterns may be significantly affected, since they are dominated by capillary flows.}

\subsubsection{Radial Velocity}\label{Radial_velocity}

The particle transport toward the droplet edge during evaporation, known as the coffee-ring effect, is driven by the radial velocity $v(r,t)$. Following Deegan et al.~\cite{Deegan2000}, it can be expressed as
\begin{equation}\label{eq:vel_r}
\begin{split}
  v(r,t) = & -\frac{1}{\rho r h} \int_0^r dr' \, r' \\
  & \times \left( J \sqrt{1 + \left(\frac{\partial h}{\partial r'}\right)^2} + \rho \frac{\partial h}{\partial t} + \zeta_{r'} \right),
\end{split}
\end{equation}
where $h(r,t)$ is the droplet interface and $J(r,t)$ the local evaporative flux. In the original formulation of Deegan~\cite{Deegan2000}, the last term $\zeta_r$ is absent; here it is introduced as a correction to account for contact line motion (see Eq.~\ref{ConservedVolume}). The relative radial velocity is defined as $\tilde{v}(r,t)=v(r,t)/v_c$.
For reaction-limited evaporation with $K \in \mathcal{O}(10)$, contact line pinning drives a radial flow since the evaporation is nearly uniform~\cite{MURISIC2011, Larsson2023}.

\begin{figure}
    \centering
    \includegraphics[width=1.0\linewidth]{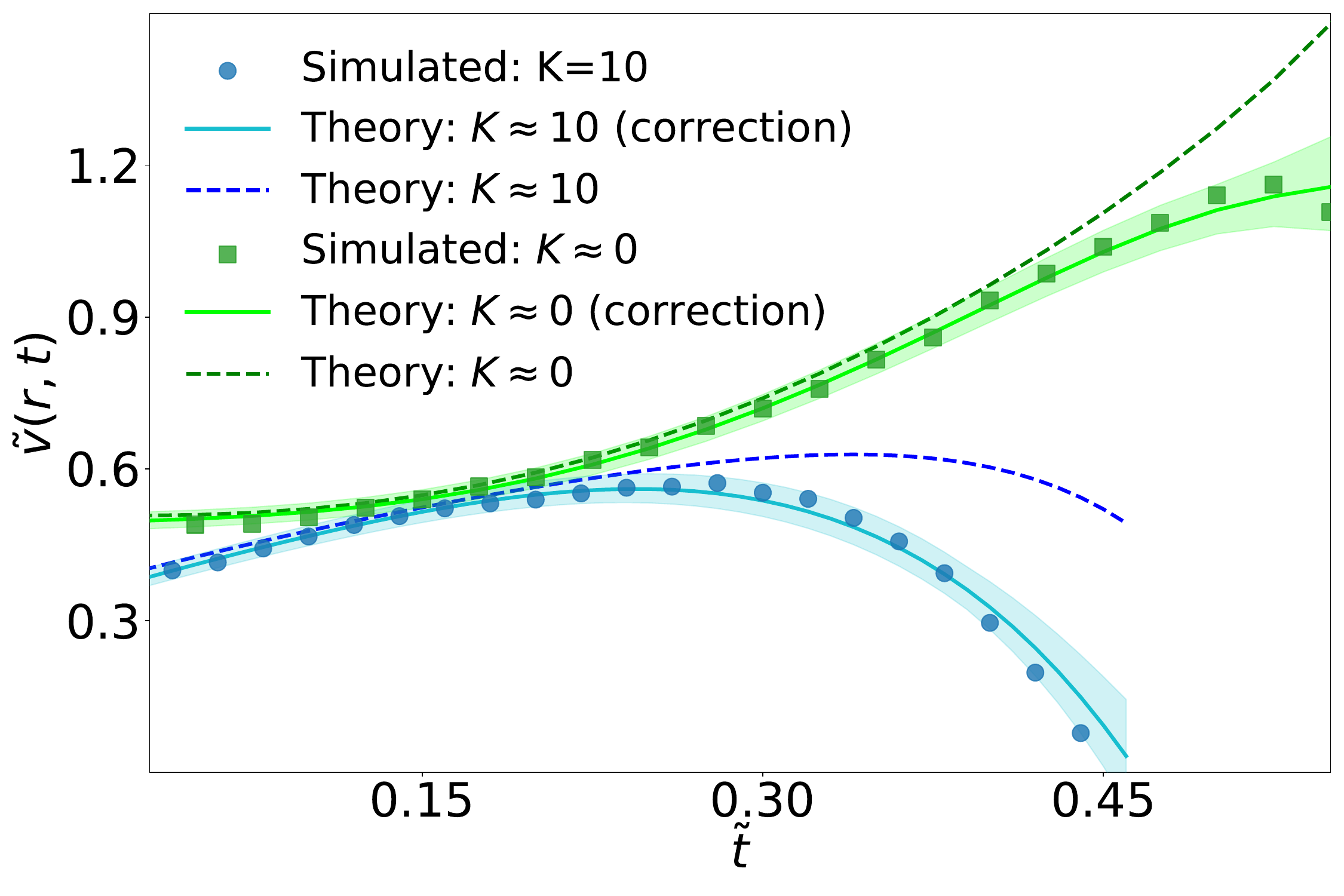}
    \caption{Time evolution $\tilde{t} =\frac{t}{N_t}$
    of the relative radial velocity $\tilde{v}(r,t)=v(r,t)/v_c$
    at $r/a=0.79$ for an evaporating droplet in stick–slide mode. Simulation results (dots) are compared to theory (Eq.~\ref{eq:vel_r}) with correction (solid lines) and without correction (dashed lines), with the time evolution of the base radius $a$ and the local height $h$ described by polynomial fits.
    }
    \label{fig:RL-RH}
\end{figure}

Eq.~\ref{eq:vel_r} without the correction term $\zeta_r$ assumes perfect contact line pinning. In contrast, in the stick-slide regime, the base radius remains nearly constant during the CCR regime, followed by a pronounced reduction in the CCA mode. Fig~\ref{fig:RL-RH} benchmarks the CGLB evaporation model against theoretical predictions with and without the correction term, considering the effect of a dynamic contact line on the radial velocity. Both the reaction-limited ($K=10$, blue) and the diffusion-limited ($K\approx0$, green) regimes show good agreement. For analysis, we exclude the initial portion corresponding to the system's equilibration.
Dashed lines denote the uncorrected theory, while solid lines include the $\zeta_r$ correction. 
Shaded regions indicate uncertainties of $\pm 0.5$ lattice units in the interface position and $1\%$ in the correction factor $S$ used in the CGLB formulation~\cite{Nath2025}.

\begin{figure*}
    \centering
    \includegraphics[width=\linewidth]{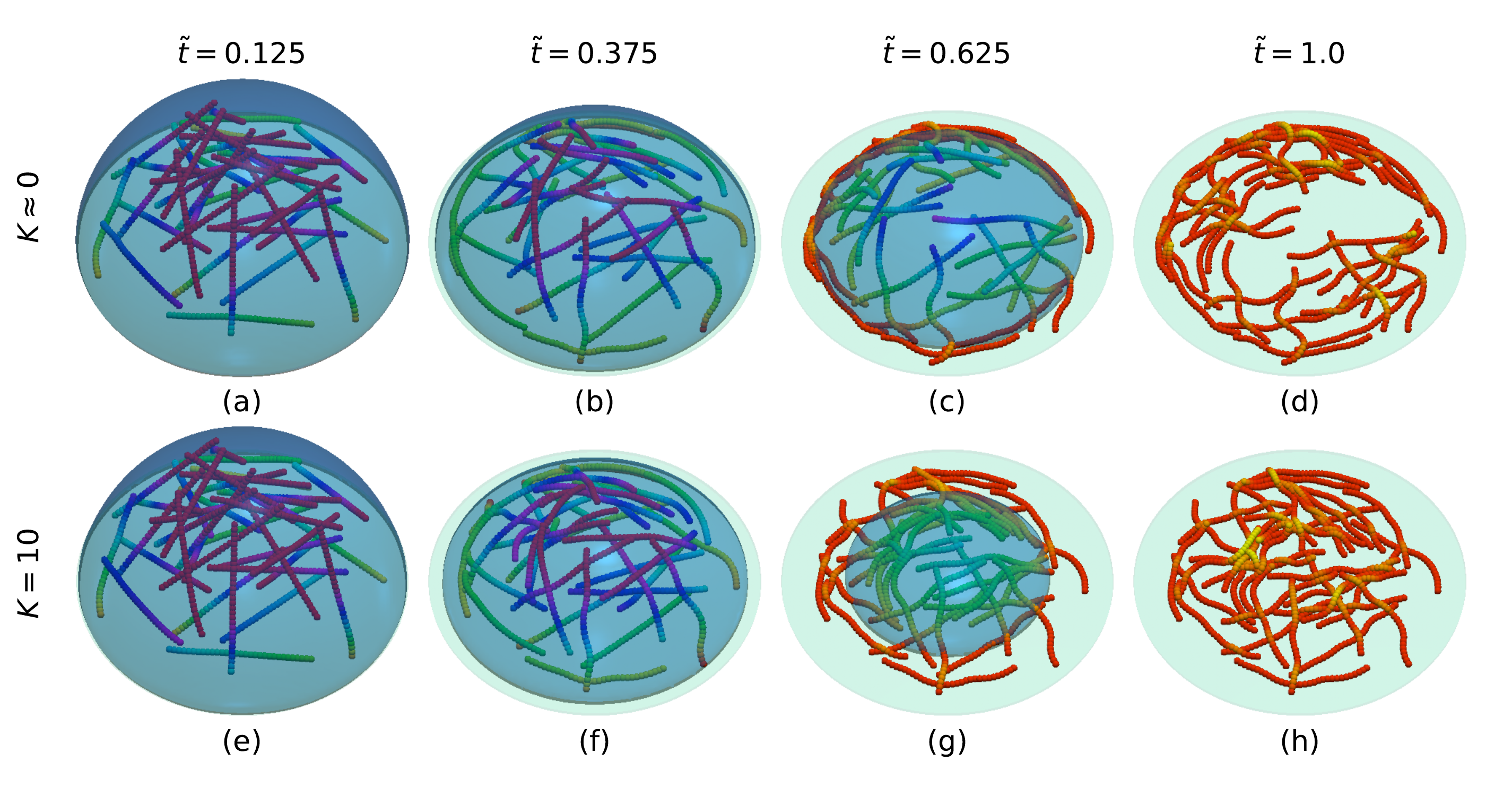}
    \caption{Snapshots of the drying process at different time $\tilde{t}$, where $\tilde{t} =\frac{t}{N_t}$ is non-dimensionalized. Top: diffusion-limited regime ($K\approx 0$). Bottom: reaction-limited regime ($K=10$). Filaments shown are 31 units long and have an area fraction of $c\approx31\%$. The color bar denotes the $z$-coordinate, with light red indicating regions nearest to the substrate and purple the most distant.}
    \label{fig:simulation_overview}
\end{figure*}

\subsection{Evaporation of a filament-laden droplet}

We now investigate the self-organization of filaments within an evaporating droplet. Fig.~\ref{fig:simulation_overview} visualizes \revisedtext{the dynamic evolution of} a filament-filled droplet drying on a substrate under reaction-limited and diffusion-limited evaporation regimes. We initialize non-overlapping straight filaments, randomly positioned and oriented within the droplet volume (Fig.~\hyperref[fig:simulation_overview]{\ref*{fig:simulation_overview}a} and \hyperref[fig:simulation_overview]{\ref*{fig:simulation_overview}e}). \revisedtext{
We note that filament flexibility is fully incorporated into our model, and the filaments are initialized in straight configurations to approximate their equilibrium state in a good solvent. Starting from straight filaments also allows us to isolate and systematically
analyze the underlying deposition mechanisms without introducing additional variability arising from arbitrary initial curvature. 
%Deposit morphology is expected to depend on pre-existing curved or highly entangled filament configurations, particularly at higher concentrations where collective effects dominate.
%We note that initializing filaments as straight objects could bias the resulting structures, as the final deposit morphology is influenced by pre-existing filament curvature, as well as stretching, bending, entanglement induced by evaporation-driven flow, filament interactions, and interface confinement.
}
During evaporation, filaments are first gradually advected toward the contact line (Fig.~\hyperref[fig:simulation_overview]{\ref*{fig:simulation_overview}b} and \hyperref[fig:simulation_overview]{\ref*{fig:simulation_overview}f}).
Bending arises primarily from solvation forces at the interface, which cause free chain segments to align with the local curvature of the droplet surface. Additional shape changes result from interparticle interactions and adhesion between deposited segments and the substrate.
In this stage, filament segments with sufficient substrate friction overcome solvation forces and remain deposited, while unconstrained segments are carried inward by the receding interface (Fig.~\hyperref[fig:simulation_overview]{\ref*{fig:simulation_overview}c} and \hyperref[fig:simulation_overview]{\ref*{fig:simulation_overview}g}).
\revisedtext{Filament segments near the contact line get effectively immobilized by friction with the substrate. When one end of a filament is anchored at the contact line, the resulting mechanical constraint is transmitted along the chain via interparticle interactions, influencing segments farther from the contact line and promoting alignment along the inward radial direction towards the droplet center.}
The filaments finally settle onto the substrate (Fig.~\hyperref[fig:simulation_overview]{\ref*{fig:simulation_overview}d} and \hyperref[fig:simulation_overview]{\ref*{fig:simulation_overview}h})~\cite{Park2024}, leading to a coffee-ring pattern in the diffusion-limited regime ($K\approx0$) and a homogeneous deposit in the reaction-limited regime ($K=10$).

\subsubsection{Order of Filament Deposits}

To characterize the structure of the filament network confined within a circular domain of radius $a$ centered at $\mathbf{r}_0 = (x_0, y_0)$, with $x_0=y_0=48$, we evaluate two metrics as functions of radial distance $r = |\mathbf{r} - \mathbf{r}_0|$:  
(i) the tangential nematic order parameter $S_{\text{t}}(r)$, measuring the local alignment of filament segments relative to the domain boundary, and  
(ii) the radial density profile $g(r)$, which quantifies the spatial distribution of filaments.

\paragraph*{Tangential nematic order parameter:}  
$S_{\text{t}}(r)$ quantifies the alignment of filament segments relative to a circular boundary, and is defined as 
\begin{equation}
    S_{\text{t}}(r) := \left\langle 2\cos^2\theta(\mathbf{r}) - 1 \right\rangle_{|\mathbf{r}-\mathbf{r}_0|=r}.
\end{equation}
Here, $\theta(\mathbf{r})$ is the angle between a segment’s orientation vector and the local tangent to the circular boundary, and $\langle \cdot \rangle$ denotes averaging over all segments at distance $r$ from $\mathbf{r}_0$.
We note that $S_{\text{t}}(r) = 1$ represents perfect tangential alignment, $S_{\text{t}}(r)= -1$ denotes radial alignment, and $S_{\text{t}}(r)\approx0$ denotes isotropic configurations.
For numerical evaluation, the system is discretized into concentric annular bins of width $\Delta r=0.1a$.
The order parameter within the $i$-th bin is computed as
\begin{equation}
    S_{\text{t}}(r_i) = \frac{1}{N_i} \sum_{j=1}^{N_i} \left[ 2\left( \mathbf{u}_j \cdot \hat{\mathbf{t}}_j \right)^2 - 1 \right],
\end{equation}
\revisedtext{where $ r_i $ denotes the radial position of the bin and $ N_i $ the number of segments within it. Here, a segment is defined as the bond connecting two consecutive monomers within a filament. The vector $ \mathbf{u}_j $ describes the local orientation of the $ j $-th segment, and $ \hat{\mathbf{t}}_j $ is the local tangent unit vector evaluated at the segment midpoint $ \mathbf{r}_j = (x_j, y_j) $:}
\begin{equation}
    \hat{\mathbf{t}}_j = \frac{(y_j - y_0,\, x_0 - x_j)}
    {\sqrt{(x_j - x_0)^2 + (y_j - y_0)^2}}
\end{equation}
This order parameter captures how circular confinement influences the particle orientation, and highlights the degree of tangential alignment induced by the circular boundary, particularly near the domain edge.

\begin{figure}
    \centering
    \includegraphics[width=1.0\linewidth]{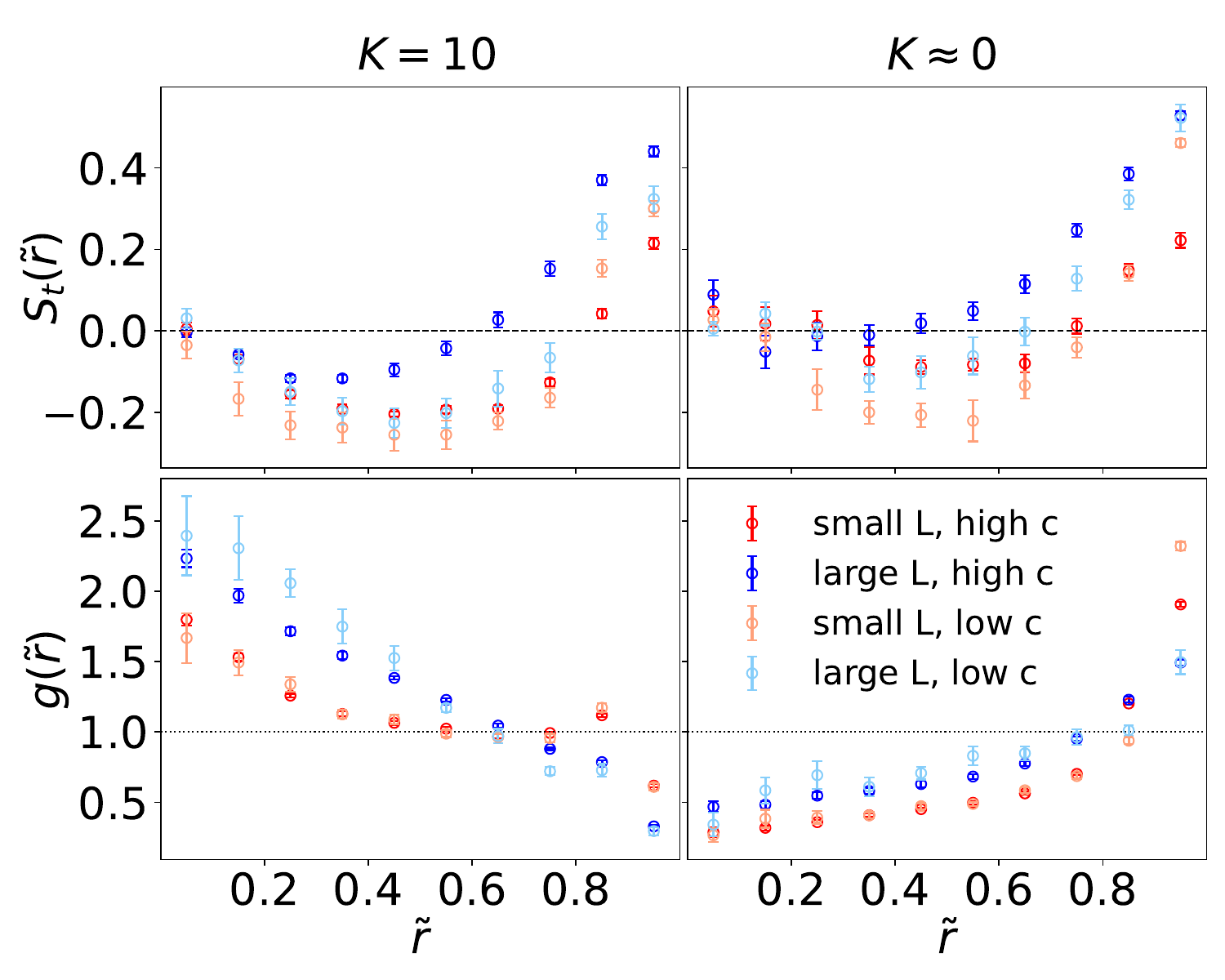}
    \caption{Nematic order $S_t(\tilde{r})$ and radial distribution $g(\tilde{r})$ versus normalized radial distance $\tilde{r}$ for fully flexible filaments in the \revisedtext{reaction}-limited evaporation $K=10$ (left) and the diffusion-limited regime $K\approx0$ (right).}
    \label{fig:rdf_nematic}
\end{figure}

\paragraph*{Radial density profile:} $g(r)$ captures the spatial distribution of beads as a function of radial distance from the center. It is defined as the ratio between the local number density $ n(r) $ and the average number density $ n_0 $ over the entire domain:
\begin{equation}
    g(r) = \frac{n(r)}{n_0}
    \label{eq:gr}
\end{equation}
Here, $n(r) = N(r) / \Delta A(r)$ is the number density within the $i$-th annular bin $[r_i, r_{i+1}]$, where $N(r)$ is the number of beads located within the bin, and $\Delta A(r) = \pi (r_{i+1}^2 - r_i^2)$ is the area of the annulus. The average density is $\revisedtext{n_0} = N_{\text{total}} / (\pi R^2)$, where $N_{\text{total}}$ is the total number of beads. Values of $g(r) > 1$ indicate enrichment, while $g(r) < 1$ signifies depletion at the corresponding radial distance.
Fig.~\ref{fig:rdf_nematic} presents $S_t(r)$ and $g(r)$ for four representative cases obtained by varying the filament length $L$ and the filament concentration $c_v$.
To improve statistical significance, data were \revisedtext{first averaged over 16 independent initial configurations for each $(L,c_v)$ parameter set and subsequently} over parameter ranges defined as “small filaments” ($L_{\mathrm{min}} \le L < L_{\mathrm{small}}$) and “large filaments” ($L_{\mathrm{large}} < L \le L_{\mathrm{max}}$), as well as “low concentrations” ($c_{\mathrm{min}} \le c_v < c_{\mathrm{low}}$) and “high concentrations” ($c_{\mathrm{high}} < c_v \le c_{\mathrm{max}}$). The corresponding thresholds were \revisedtext{$L_{\mathrm{min}}=3$, }$L_{\mathrm{small}}=11$, $L_{\mathrm{large}}=23$, $L_{\mathrm{max}}=31$, \revisedtext{$c_{\mathrm{min}}=0.0012$, }$c_{\mathrm{low}}=0.0035$, $c_{\mathrm{high}}=0.0107$, and $c_{\mathrm{max}}\approx 0.0131$. The averaging was performed with step sizes of $\Delta L = 2$ and $\Delta c_v = 0.0012$. \revisedtext{Fig.~S1 and S2 in the Supporting Information show results that are averaged only over different initial configurations, but not over different $(L,c_v)$ parameter values. Within these subsets, no systematic trends are observed.}
Three characteristic radial regions can be distinguished. As shown in Fig.~\ref{fig:rdf_nematic}, in the \textit{outer region} ($\tilde{r} \gtrsim 0.75$), where $\tilde{r}=r/a_{max}$ and $a_{max}=32.79$ denotes the average maximal deposit radius, we find $S_t(r) > 0$, indicating pronounced tangential alignment (corresponding to Fig.~\hyperref[fig:simulation_overview]{\ref*{fig:simulation_overview}d} and \hyperref[fig:simulation_overview]{\ref*{fig:simulation_overview}h}). This alignment results from outward capillary flows during the CCR phase, with filaments advected to the interface adapting to its curvature. In the \textit{intermediate region} ($0.25 \lesssim \tilde{r} < 0.75$), $S_t(r) < 0$ reflects radial alignment. Here, shear gradients during the CCR phase stretch filaments radially, while in the CCA phase the receding contact line reinforces this orientation as anchored segments resist the inward motion of free segments. In the \textit{inner region} ($\tilde{r} \lesssim 0.25$), high particle density or low radial flow and excluded-volume effects suppress radial alignment, leading to weak tangential ordering as filaments aggregate into an approximately isotropic central cluster.

\revisedtext{The magnitude and spatial extent of these alignment patterns depend on filament length: longer filaments show stronger tangential alignment, particularly outside the inner region, which can be attributed to their lower effective number of inter-particle contacts at the same filament concentration. In contrast, shorter filaments experience more frequent interactions that disrupt orientational order (top row of Fig.~\ref{fig:rdf_nematic}).}

Evaporation conditions further modulate this behavior. In the diffusion-limited regime ($K \approx 0$), strong capillary flows persist into the CCA phase (see Fig.~\ref{fig:RL-RH}), transporting particles toward the contact line and enhancing tangential alignment (see Fig.~\ref{fig:rdf_nematic}).
In contrast, in the reaction-limited regime ($K = 10$), uniform evaporation suppresses capillary flows and inward contact line motion dominates, resulting in weaker tangential order. \revisedtext{For short filaments, the difference is negligible at low concentrations, where inter-particle interactions are reduced.}

The radial density profile $g(r)$ supports these trends. 
For $K=10$, central enrichment and edge depletion are observed: short filaments exhibit a weak coffee-ring peak near $\tilde{r} \approx 0.85$, whereas long filaments remain concentrated near the center and show no distinct coffee-ring peak. Under $K \approx 0$, central depletion and edge accumulation occur, yet long filaments still exhibit a \revisedtext{more homogeneous distribution.}

\begin{figure*}
    \centering
    \includegraphics[width=1\linewidth]{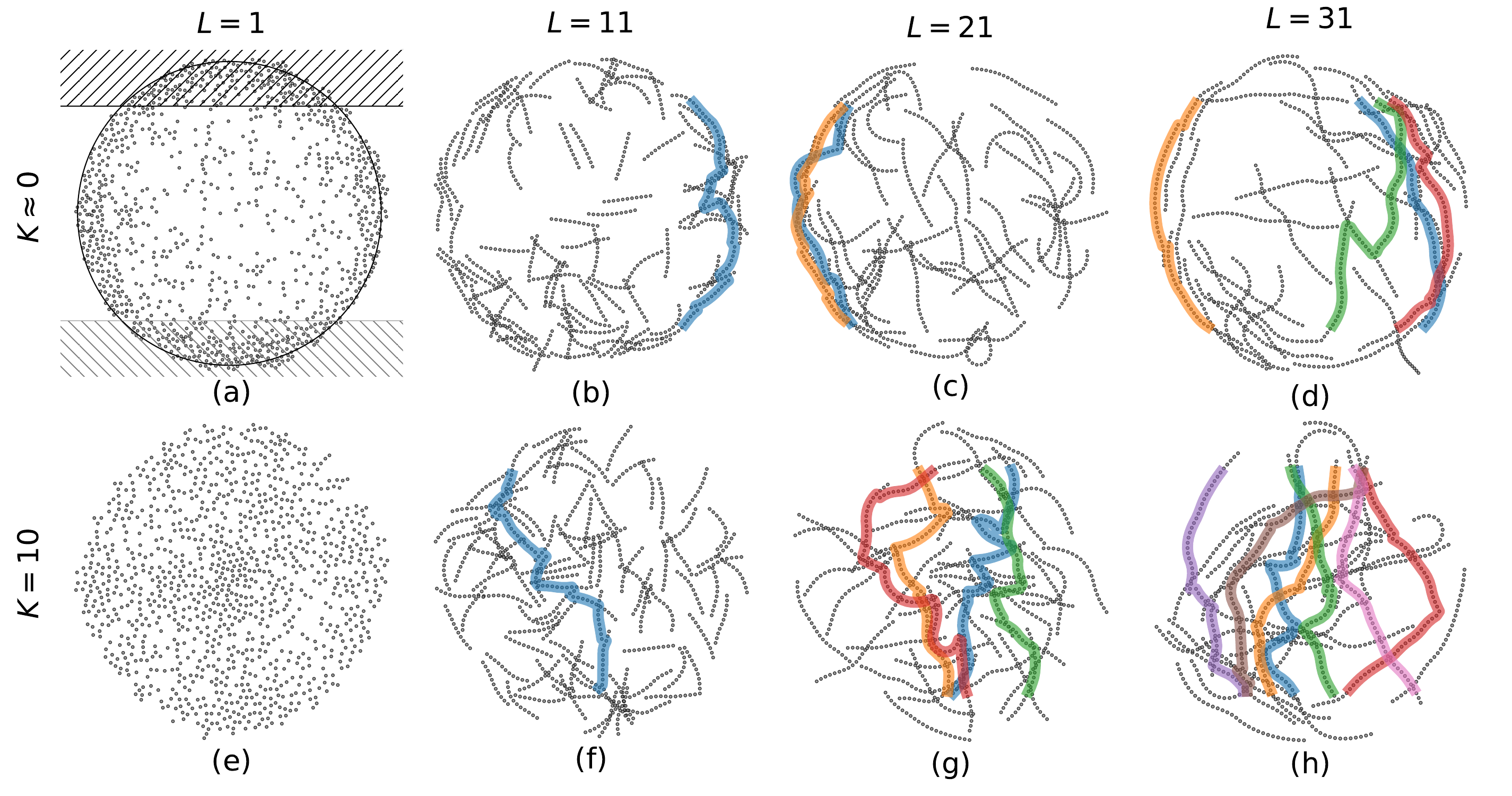}
    \caption{Deposition patterns for different filament lengths (left to right) 1, 11, 21, and 31 at a given area fraction $c \approx 31\%$.
    Top: diffusion-limited evaporation ($K \approx 0$). Bottom: reaction-limited evaporation ($K = 10$). Colored lines indicate distinct, independent percolation pathways. In subfigure (a) the definition of source and sink is illustrated.}
    \label{fig:PercolativePathways2}
\end{figure*}

\subsubsection{Percolation of filament deposits}\label{S:Perc}

Percolation theory describes the emergence of long-range connectivity in disordered systems and underpins transport in polymeric, colloidal, and nanowire networks~\cite{Kirkpatrick1973}. We investigate a two-dimensional network of filaments confined within a circular domain of radius $a$, where chains of length $L$ (bead diameter $d_0$) are deposited. Two beads are defined as connected if their center-to-center distance is less than $d_c = 1.12 d_0$.

We investigate filament percolation in layers deposited from evaporating droplets. To evaluate percolation, we define source and sink regions as two spherical-cap sectors located on opposite sides of the circular simulation domain, with a distance of $0.8a$.
\revisedtext{To compare to percolation in the absence of} hydrodynamic alignment and curvature, we generated an isotropic control system via Monte Carlo (MC) insertion of straight filaments into a periodic rectangular box matching the caps' separation and base diameter.

A percolation pathway is a connected sequence of beads that spans from the source to the sink. We define independent percolation pathways (IPPs) as disjoint sets of such paths. For each configuration, we determine the number of IPPs, $N_{\mathrm{IPP}}$, and compute the ensemble average $\langle N_{\mathrm{IPP}} \rangle$ over 16 random initializations. Fig.~\ref{fig:PercolativePathways2} illustrates IPPs within the deposit configurations for different filament lengths.
For short filaments, no IPPs are observed, as shown in Fig.~\hyperref[fig:PercolativePathways2]{\ref*{fig:PercolativePathways2}a} and \hyperref[fig:PercolativePathways2]{\ref*{fig:PercolativePathways2}e}. With increasing length of the filaments the number of IPPs increases on average, as shown for example in Fig.~\hyperref[fig:PercolativePathways2]{\ref*{fig:PercolativePathways2}d} and \hyperref[fig:PercolativePathways2]{\ref*{fig:PercolativePathways2}h}. In the reaction-limited regime ($K=10$) we observe on average an increased number of IPPs. 
For lower critical contact angles, a stronger CRE is expected, especially for the diffusion-limited regime ($K\approx0$), which would further increase IPP formation along the perimeter.

The \textit{percolation threshold} $n_c$ is the critical filament number density at which a spanning cluster first appears. It is obtained as the inflection point of a sigmoid fit to the percolation probability $P(n)$ as a function of filament number density $n$~\cite{Pawowska2013}:  
\begin{equation}
    P(n) = \left[ 1 + \exp\left( -\frac{n - n_c}{w} \right) \right]^{-1}
\end{equation}
Here, $w$ denotes the transition width.
\revisedtext{For filaments of finite aspect ratio}, the critical number density $n_c$ decreases with increasing aspect ratio, in agreement with excluded volume theory~\cite{Balberg1984, Broman2021, Singh2025}:
\begin{equation}\label{crit_number}
    n_c = C\,L^{-p} \revisedtext{+ C_0},
\end{equation}
where $C$ is a fitting parameter. \revisedtext{$p=2$ is a theoretical value for the case of high aspect-ratio filaments~\cite{Balberg1984, Mietta2014, Li2009} and $C_0$ accounts for finite-size effects outside of the transient regime for high aspect ratios.}
\begin{figure*}
    \centering
    \includegraphics[width=\textwidth]{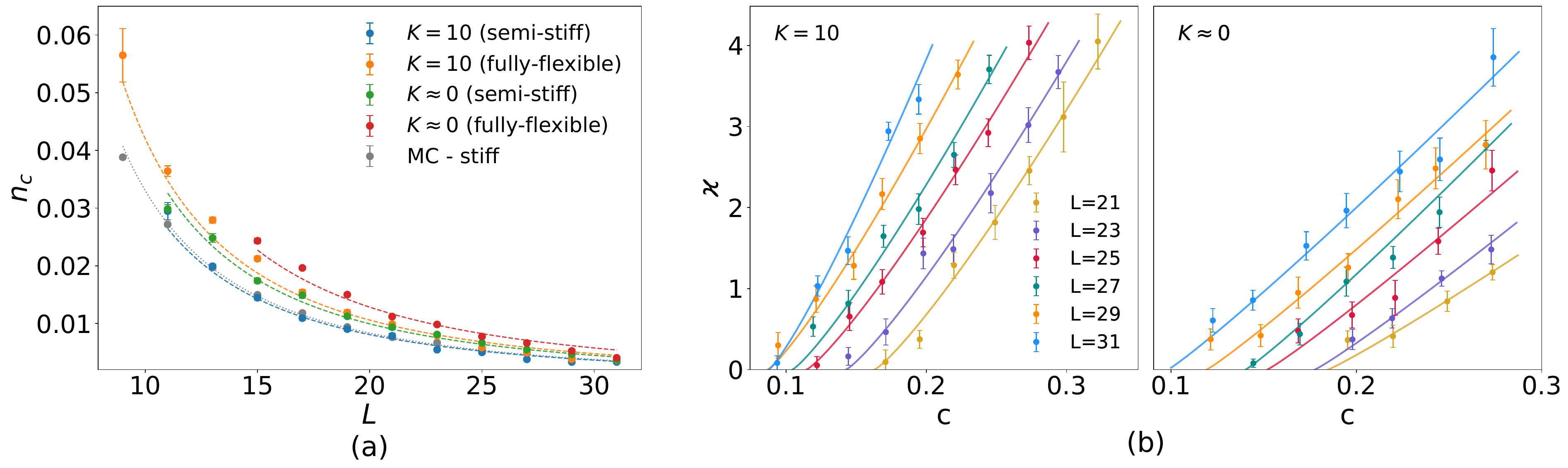}
    \caption{The distance between source and sink is chosen to be $0.8a$. (a) Percolation threshold versus filament length for fully flexible and semi-stiff filaments under reaction-limited ($K=10$) and diffusion-limited ($K \approx 0$) evaporation, compared with Monte Carlo data. We observe for stiffer filaments and more uniform evaporation a decrease in the percolation thresholds. (b) Conductivity $\varkappa$ vs. area fraction $c$ for fully-flexible filaments in both evaporation regimes: $K=10$ (left) and $K \approx 0$ (right). We observe a larger critical exponent for more uniform evaporation.}
    \label{Perc_cond}
\end{figure*}

Fig.~\hyperref[Perc_cond]{\ref*{Perc_cond}a} shows the percolation thresholds as a function of filament length for a source–sink distance of $0.8a$. \revisedtext{For short filaments, percolation events become increasingly rare and could therefore not be sampled with sufficient statistical accuracy.}
\revisedtext{Results for fully flexible and semi-stiff filaments under reaction- and diffusion-limited evaporation are fitted using Eq.~\ref{crit_number}, with the finite-size bias $C_0=1 \times 10^{-4}$ determined from the Monte Carlo data and fixed for the remaining parameters. The fit parameters are listed in Tab.~\ref{tab:ab_parameters}.} 
For filament lengths exceeding 10, the semi-stiff case with $K=10$ yields percolation thresholds comparable to the Monte Carlo reference, as the resulting deposits consist of relatively straight, uniformly distributed filaments similar to those in the random placement model.
\revisedtext{In contrast, all other cases exhibit higher thresholds, either due to the reduced effective length of coiled, flexible filaments or due to radially inhomogeneous deposits that lie between a homogeneous film and a fully developed coffee-ring structure. In such morphologies, network connectivity is reduced because the central region becomes depleted before the peripheral accumulation is sufficiently dense to establish percolation, leading to a dip in conductivity.}

\begin{table}[htbp]
\centering
\begin{tabular}{lcccc}
\textbf{Condition} & $\boldsymbol{C \pm \delta C}$ \\
\hline
$K \approx 0$ (fully-flex)   & $5.09 \pm 0.18$ \\
$K \approx 0$ (semi-stiff)   & $3.93 \pm 0.07$ \\
$K = 10$ (fully-flex)        & $4.19 \pm 0.16$ \\
$K = 10$ (semi-stiff)        & $3.19 \pm 0.07$ \\
MC – stiff                   & $3.29 \pm 0.03$ \\
\end{tabular}
\caption{\revisedtext{
Fitted values of the parameter \( C \) with standard errors from Eq.~\ref{crit_number}, which models the percolation threshold as a function of filament length. The finite size bias \( C_0 = 1 \times 10^{-4} \) was determined for the MC data and fixed for the other parameters. The results indicate that stiffer filaments and uniform evaporation conditions yield to lower percolation thresholds.}}
\label{tab:ab_parameters}
\end{table}

\subsubsection{Conductivity of the Deposit}\label{S:Cond}
Near the percolation threshold, the conductivity as a function of concentration is expected to follow a power-law scaling~\cite{Kirkpatrick1973}:
\begin{equation}\label{Pwlaw}
    \varkappa(c) \propto (c - c_t)^\alpha \quad \text{for } c > c_t,
\end{equation}
where $c = nLW$ is the area fraction, $\alpha$ the critical conductivity exponent, and $c_t := n_cLW$ the critical area fraction. Here, $W\revisedtext{=2d_0}$ is the filament width. In the thermodynamic limit, this scaling arises from the scale-invariant nature of the system at the threshold~\cite{Christensen2005}. While our finite systems do not exhibit strict fractality, we expect an effective scaling behavior consistent with percolation theory.

We treat the deposited filaments as a resistor network and compute the effective conductivity by applying Kirchhoff’s laws~\cite{Kirkpatrick1973}, normalizing by the source–sink distance $0.8a$.
To mitigate finite-size bias, we define the background level $\varkappa_{\mathrm{bg}}$ as the conductivity observed where the percolation probability reaches $50\%$\revisedtext{~\cite{stauffer1992}}. The effective conductivity is then obtained by subtracting this baseline: $\varkappa = \varkappa_{\mathrm{raw}} - \varkappa_{\mathrm{bg}}$.

To determine the scaling behavior of conductivity near the percolation threshold, we performed a simultaneous nonlinear regression across multiple datasets. Each dataset corresponds to a different filament length $L$ and contains measurements of $\varkappa$ as a function of $c$. The data are fitted to the power-law form:
\begin{equation}
  \varkappa(c) = a_L \,[c - c_t^{(L)}]^\alpha,  
\end{equation}
where $a_L$ is a length-dependent prefactor and $c_t^{(L)}$ is the fixed critical concentration for each $L$.
Only data points just above $c_t^{(L)}$, where conductivities remain low and scaling behavior is expected, are included.
Model parameters are obtained by minimizing the chi-squared statistic, i.e., the total weighted sum of squared residuals: 
\begin{equation}
    \chi^2 = \sum_L \sum_i \left[ \frac{\varkappa_i^{(L)} - a_L (c_i^{(L)} - c_t^{(L)})^\alpha}{\delta \varkappa_i^{(L)}} \right]^2,
\end{equation}
where $\varkappa_i^{(L)}$ is the measured conductivity, $\delta \varkappa_i^{(L)}$ is its uncertainty, and the sums run over all data points and filament lengths. Minimization of $\chi^2$ provides the best-fit parameters, with the shared exponent $\alpha$ and the prefactors $a_L$ determined simultaneously. Parameter uncertainties are estimated from the local curvature of $\chi^2$ near its minimum, using a numerical approximation to the Hessian matrix.

\revisedtext{Given the limited dynamic range, we adopt a power-law ansatz motivated by percolation theory~\cite{Clauset2009, Sahimi2023} and focus on extracting the scaling exponent.
Fig.~\hyperref[Perc_cond]{\ref*{Perc_cond}b} reports the fully flexible case under reaction- and diffusion-limited evaporation, fitted to the background-corrected data to extract 
the shared conductivity exponent $\alpha$ near $c_t$ (shown in Tab.~\ref{tab:flexibility_K_valuesGYR}).}
\revisedtext{These exponents deviate from the universal Monte Carlo reference value and exhibit weak trends with evaporation regime and filament stiffness, with partially overlapping confidence intervals. The largest values occur for uniform evaporation in the reaction-limited regime, particularly for fully flexible filaments.} \revisedtext{These exponents deviate from the universal Monte Carlo value and show weak trends with evaporation regime and filament stiffness. 
The deviation likely arises from microscopic network effects: filament flexibility and evaporation-driven aggregation create local clustering and correlations, modifying connectivity and slightly shifting the effective conductivity exponent compared to ideal percolation.}

\begin{figure*}[t!]
    \centering
    \subfloat[Filament \revisedtext{area fraction} $c$. Left: $c=0.156$, right: $c=0.219$. \label{fig:ad12}]{
        \includegraphics[width=0.6\textwidth]{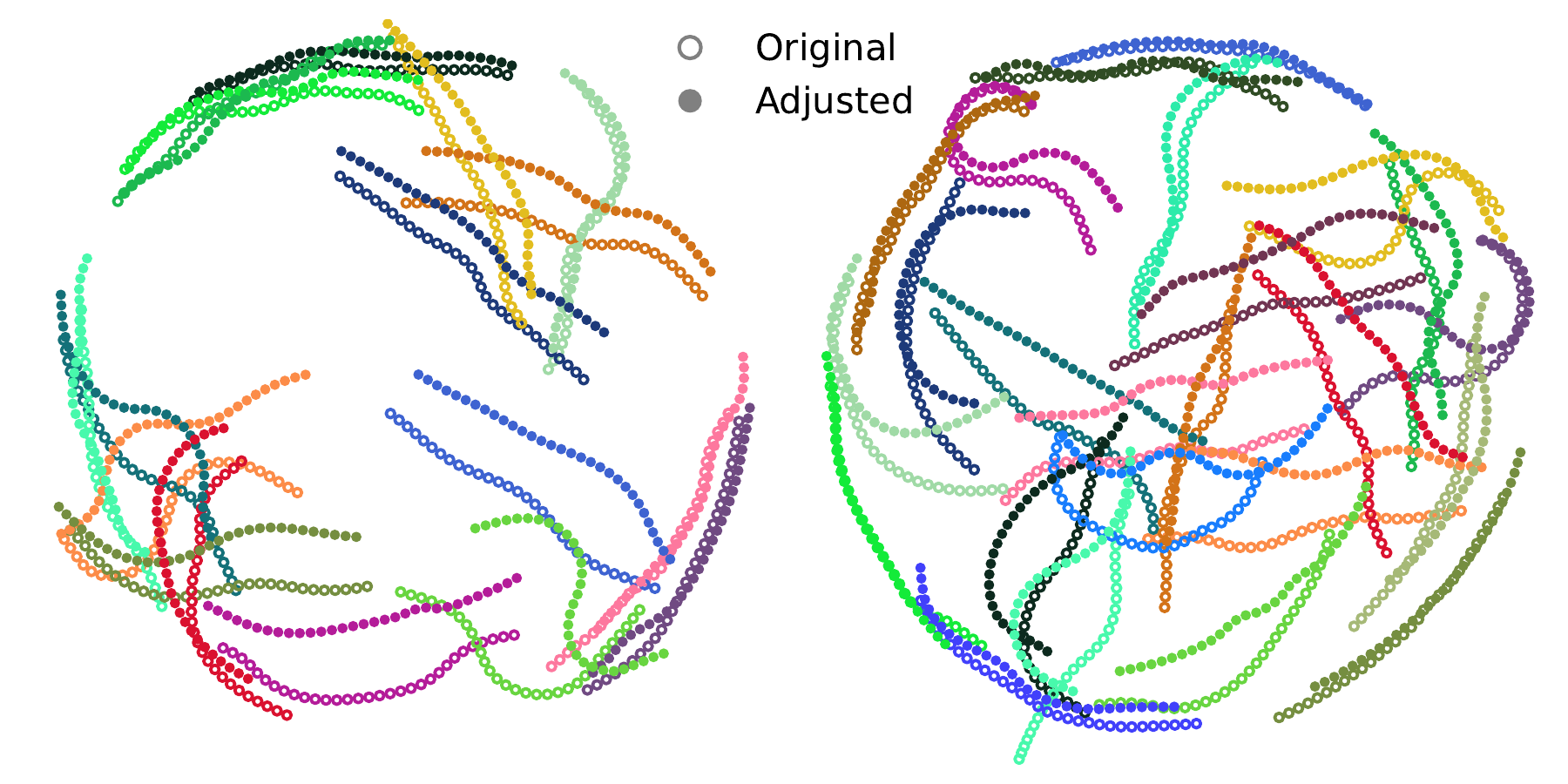}
    }
    \hfill
    \subfloat[\revisedtext{Angle-resolved n}ormalized conductivity and \revisedtext{area fraction}. \label{fig:ad3}]{
        \includegraphics[width=0.33\textwidth]{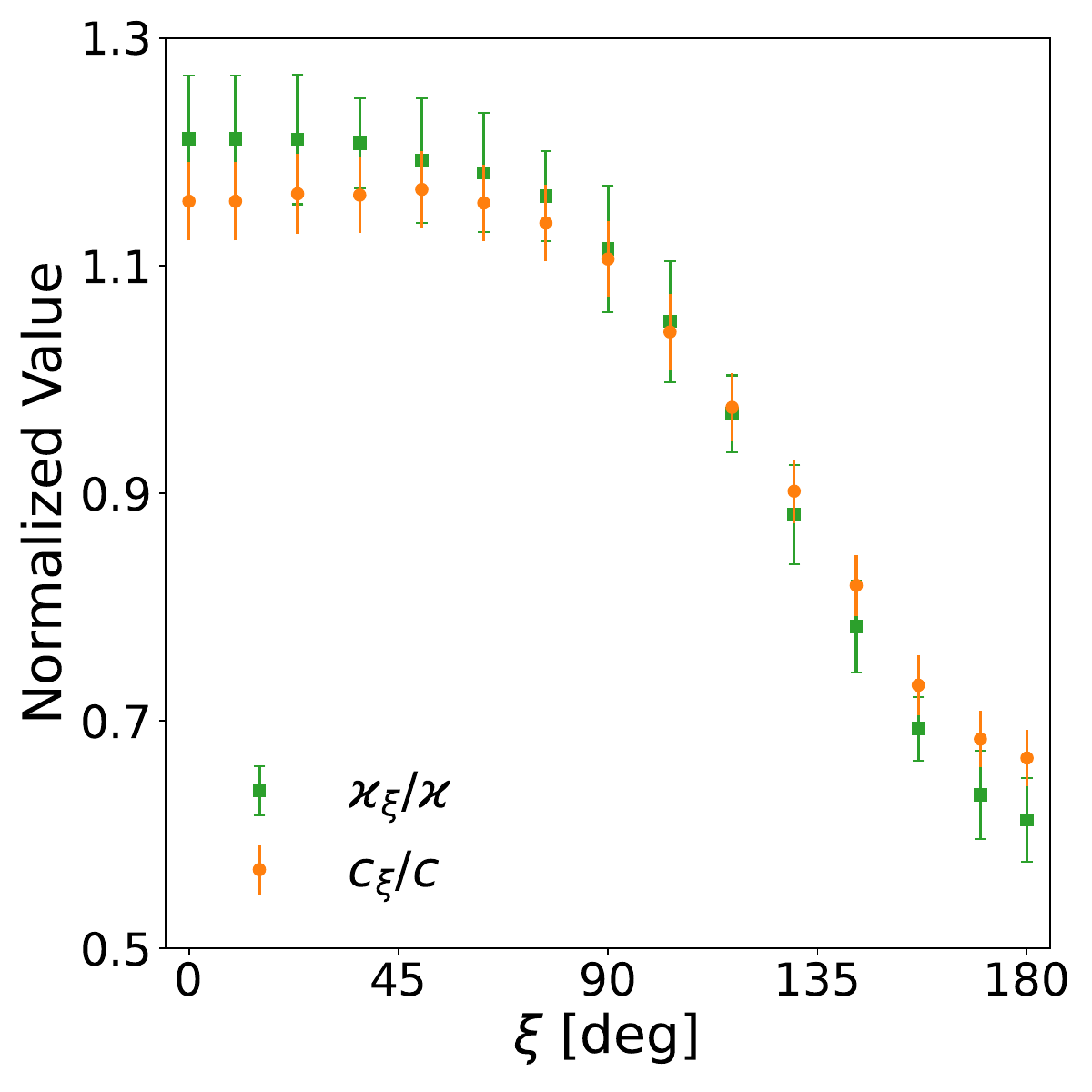}
    }
    \caption{Snapshot of deposition pattern with semi-stiff filaments at length $L = 31$ under diffusion-limited evaporation ($K \approx 0$), illustrating the vapor-shielding effect between neighbouring droplets. (a) Deposition pattern of a single isolated droplet on a substrate (open circles) compared to a droplet in the presence of a neighbor (filled circles) for different area fractions $c=0.156$ (left) and $c=0.219$ (right). The neighboring droplet is located directly below at a center-to-center distance $b$ relative to the base radius $a$, with a ratio of $a/b=0.002$.
    (b) Smoothed normalized conductivity $\varkappa_{\xi}/\varkappa$ (squares) and area fraction $c_{\xi}/c$ (circles) averaged over the left and right half-space of the circular domain, plotted versus the azimuthal angle from the droplet center. Conductivity is evaluated perpendicular to this angle and averaged over eight independent initial configurations per area fraction ($c = 0.156,\, 0.219,\, 0.282,\, 0.345$). $\xi=180^\circ$ corresponds to the droplet–droplet axis. \revisedtext{Both quantities are normalized by their respective angle-averaged values.}}
    \label{fig:adComb}
\end{figure*}
\subsubsection{Two closely adjacent droplets}

A reference Monte Carlo simulation in the same domain yields $\alpha = 1.26 \pm 0.01$, in agreement with the universal two-dimensional conductivity exponent for random percolation ($\alpha \approx 1.3$)~\cite{Berg2024} and with continuum stick-network simulations reporting $\alpha = 1.24 \pm 0.03$~\cite{Balberg1983} and $\alpha = 1.280 \pm 0.014$~\cite{Li2010}.

\begin{table}[htbp]
    \centering
    \begin{tabular}{|c|c|c|}
        \hline
        & Fully flexible & Semi-stiff \\
        \hline
        $K \approx 0$ & $1.081 \pm 0.040$ & $1.094 \pm 0.026$ \\
        \hline
        $K = 10$ & $1.183 \pm 0.028$ & $1.126 \pm 0.037$ \\
        \hline
    \end{tabular}
    \caption{Conductivity exponent $\alpha$ for a source–sink distance of $0.8a$. The reference MC-simulation yields $\alpha=1.260\pm0.006$}
    \label{tab:flexibility_K_valuesGYR}
\end{table}

In the diffusion-limited regime ($K \approx 0$), the droplet-evaporation flux is highly sensitive to neighboring droplets, nearby obstacles, and external aerodynamic conditions~\cite{Zhang2025, Wang2022}. The evaporation profiles for interacting droplets are described by Eq.~\ref{eq:local_flux}. The CGLB method is well-suited to efficiently resolve the fluid dynamics in these coupled systems, where analytical solutions of the evaporation profile are known.
Fig.~\ref{fig:adComb} illustrates how a neighboring droplet modifies the deposition pattern. 
In Fig.~\ref{fig:ad12}, filament deposits from an isolated droplet (open circles) are compared with those from a droplet adjacent below (filled circles). 
For the isolated droplet, deposition is on average isotropic, reflecting symmetric capillary flows. 
With an adjacent droplet, vapor shielding reduces evaporation on the facing side, leading to fewer deposited filaments and a shift of deposits toward the non-shielded side.
Fig.~\ref{fig:ad3} presents the smoothed normalized conductivity and area fraction (circular moving average, window size 6)\revisedtext{, where both quantities are normalized by their respective angle-averaged values over the full circular domain to highlight relative angular variations, as functions of the azimuthal angle. W}here $180^\circ$ denotes the direction toward the neighboring droplet.
Conductivity is evaluated only within the half-space corresponding to each angle to resolve anisotropy; for example, at $180^\circ$, only the half-space facing the neighboring droplet is included. \revisedtext{This definition neglects percolating pathways that cross between half-spaces and therefore slightly underestimates the absolute conductivity; however, it is justified here as a measure of the angle-dependent average value.}
Consequently, both conductivity and area fraction reach minima at $180^\circ$, dropping by approximately $50\%$ and $40\%$ respectively relative to the unshielded side at $0^\circ$. This anisotropy is tunable via inter-droplet spacing: reducing the separation intensifies vapor shielding, thereby amplifying the contrast between facing and outer boundaries.
\revisedtext{We propose that the angular anisotropy in conductivity and filament area fraction can serve as an experimental indicator of the evaporation regime, as its magnitude directly reflects the strength of vapor shielding and depends on inter-droplet spacing.}

\section{Conclusion}

We developed and validated a computational framework for simulating the evaporation-driven deposition of sessile droplets containing filaments on a substrate. The framework couples the color-gradient lattice Boltzmann method for multicomponent fluid dynamics with a point-particle approach for filaments. Filaments are represented by a bead–spring chain with tunable stiffness, enabling control over their flexibility and persistence length. The model includes two-way fluid–filament coupling, evaporation in reaction- and diffusion-limited regimes, solvation forces, excluded volume interactions, filament stiffness, and substrate adhesion from which frictional forces are derived.

Our method was validated by simulating pure (filament-free) droplet evaporation under the spherical cap approximation in both reaction- and diffusion-limited evaporation regimes. The simulations reproduce the expected temporal evolution of droplet height and internal velocity fields. 
Using this framework, we examined self-organization of the filament and final deposition patterns \revisedtext{for dilute filament suspensions} in the stick–slip contact line mode for medium contact angles ($\theta\approx40^\circ)$, strong substrate friction\revisedtext{, and limited filament-droplet size ratios}. Reaction-limited evaporation yields uniform, spatially extended deposits, favorable for continuous conductive pathways in, e.g., flexible, transparent, or radio-frequency electronics. In contrast, diffusion-limited evaporation leads to pronounced coffee-ring formation. The onset of the coffee-ring formation coincides with reduced network connectivity.
The orientation of deposited filaments varies across the droplet: tangential alignment dominates near the contact line, radial alignment emerges in the intermediate region, and orientations are nearly random at the center. Longer filaments preferentially align tangentially and yield more centralized deposits. The percolation threshold decreases with increasing filament length, stiffness, and with more spatially uniform evaporation (e.g., in the reaction-limited regime) \revisedtext{within the concentration range considered here}.
The resulting deposit conductivity is well described by a fitted power-law dependence on filament concentration, with modest deviations from the expected universal scaling that depend on filament stiffness and the evaporation regime.
These results are compared with Monte Carlo simulations of randomly oriented filament systems.
In diffusion-limited systems of closely spaced droplets, the facing sides exhibit reduced conductivity and a more uniform but lower-density deposition, resulting from suppressed radial flow due to reduced evaporation caused by elevated vapor concentrations between the droplets. We propose this configuration for the experimental validation of the evaporation regime.

A recent study reported a link between morphology scaling and conductivity scaling~\cite{Berg2024}. Future work should test this relationship for different evaporation-driven deposits and further investigate critical behavior~\cite{Balberg1983}. Junction resistance, determined by the number and quality of inter-filament contacts, should also be considered as it can substantially affect the effective conductivity~\cite{DeStefano2019}. \revisedtext{Additionally, it is important to investigate how filament entanglement and variations in the receding contact angle affect the deposited morphology. Higher-resolution analysis would also enable a more detailed study of filament alignment near the contact line.}
Beyond sessile droplets, our framework can be applied to multilayer deposits and line-shaped droplet geometries that are central to printing and coating technologies.

\section*{Acknowledgments}
We thank Johannes Hielscher for supporting the development of the simulation code. 
We acknowledge financial support from the Deutsche
Forschungsgemeinschaft (DFG, German Research Foundation) -- Project-ID 528402728 (research group ``3D-HF-MID") 
and the German Federal Ministry of Education and Research (BMBF) -- Project H2Giga/AEM-Direkt (Grant number 03HY103HF), as well as the project ``H2Season"
supported by the Bavarian Ministry for Economy, State Development and Energy.
We thank the Gauss Centre for Supercomputing e.V.(\url{www.gauss-centre.eu}) for funding this project by providing computing time
through the John von Neumann Institute for Computing (NIC) on the GCS Supercomputer JUWELS at Jülich Supercomputing Centre (JSC).

\section*{Declaration of competing interest}
The authors declare that they have no known competing financial interests or personal relationships that could have appeared to influence the work reported in this paper.

\section*{Data availability}
The data that support the findings of this study are openly
available at \url{https://doi.org/10.5281/zenodo.17912741}.

\bibliography{Bibfile}

\end{document}